\documentclass[aps, prd, superscriptaddress,groupedaddress,nofootinbib]{revtex4}  %

\usepackage{ifpdf}
\usepackage{graphicx}  
\usepackage{subcaption}
\usepackage{dcolumn}   
\usepackage{bm}        
\usepackage{amssymb}   
\usepackage{bbm}        
\usepackage{amsfonts, pifont}

\usepackage{amsmath,amssymb}
\usepackage{hyperref}
\usepackage{verbatim}
\usepackage{float}
\usepackage{subcaption} 
\usepackage{amsfonts}
\usepackage[inline]{enumitem}
\newlist{inlinelist}{enumerate*}{1}
\setlist*[inlinelist,1]{%
  label=(\arabic*),
}
\usepackage[normalem]{ulem}
\usepackage[usenames,dvipsnames]{xcolor}
\definecolor{darkerblue}{rgb}{0.2,0.2,0.5}

\usepackage{afterpage}

\renewcommand{\arraystretch}{1.25}
\newcommand{\xmark}{\ding{55}}

\newcommand{\bear}{\begin{array}}
\newcommand{\ear}{\end{array}}

\newcommand{\beq}{\begin{eqnarray}}
\newcommand{\eeq}{\end{eqnarray}}
\newcommand{\beqa}{\begin{eqnarray}}
\newcommand{\eeqa}{\end{eqnarray}}
\newcommand{\no}{\nonumber}

\def\OMIT#1{{}}
\newcommand{\lsim}{\mathrel{\rlap{\lower4pt\hbox{\hskip1pt$\sim$}}
    \raise1pt\hbox{$<$}}}         
\newcommand{\gsim}{\mathrel{\rlap{\lower4pt\hbox{\hskip1pt$\sim$}}
    \raise1pt\hbox{$>$}}}         

\newcommand{\dd}{\mathrm{d}}

\newcommand{\FF}{\mathcal{F}}

\newcommand{\OO}{\mathcal{O}}

\newcommand{\Gaia}{\textit{Gaia} \!}

\usepackage{diagbox}
\usepackage[flushleft]{threeparttable}
\usepackage{multirow}
\usepackage{mathtools}
\newcommand{\loge}{\text{ln}}

\makeatletter
\newcommand*{\rom}[1]{\expandafter\@slowromancap\romannumeral #1@}
\usepackage{pbox}

\begin{document}

\title{Implications of the {\it Gaia} Sausage for Dark Matter Nuclear Interactions}
\author{Jatan Buch} \email{jatan\textunderscore buch@brown.edu} 
\author{JiJi Fan} \email{jiji\textunderscore fan@brown.edu}
\author{John Shing Chau Leung} \email{shing\textunderscore chau\textunderscore leung@brown.edu}
\affiliation{Department of Physics, Brown University, Providence, RI, 02912, USA}
\date{\today}

\begin{abstract}
The advent of the {\it Gaia} era has led to potentially revolutionary understanding of dark matter (DM) dynamics in our galaxy, which has important consequences for direct detection (DD) experiments. In this paper, we study how the empirical DM velocity distribution inferred from {\it Gaia}-Sausage, a dominant substructure in the solar neighborhood, affects the interpretation of DD data. We survey different classes of operators in the non-relativistic effective field theory that could arise from several relativistic benchmark models and emphasize that the {\it Gaia} velocity distribution could modify both the total number of events as well as the shape of the differential recoil spectra, the two primary observables in DD experiments. Employing the euclideanized signal method, we investigate the effects of the {\it Gaia} distribution on reconstructing DM model parameters and identifying the correct DM model given a positive signal at future DD experiments. We find that for light DM with mass $\sim 10$ GeV, the {\it Gaia} distribution poses an additional challenge for characterizing DM interactions with ordinary matter, which may be addressed by combining complementary DD experiments with different targets and lowering the detection threshold. Meanwhile, for heavy DM with mass above ${\sim 30}$ GeV, depending on the type of DM model, there could be a (moderate) improvement in the sensitivity at future DD experiments. 

\end{abstract}

\maketitle

\section{Introduction}
\label{sec:introduction}

Confirming the existence of dark matter (DM) through a variety of cosmological and astrophysical observations has been one of the major successes of $20^{\rm th}$ century physics. Simultaneously, questions regarding the particle nature of DM and its interactions with ordinary matter beyond gravity remain unresolved. Fortunately, there is a vibrant research program that seeks to answer these questions on the experimental and observational frontiers. A leading probe in the hunt for DM is direct detection (DD) experiments, which look for signals from DM particles scattering in underground detectors. Although there have been no statistically significant detections of non-background events so far, next-generation experiments such as \textsc{Lux-Zeplin} (LZ)~\cite{Akerib:2018lyp}, \textsc{Xenon-nT}~\cite{Plante2016}, \textsc{PandaX-xT}~\cite{Zhang:2018xdp}, SuperCDMS SNOLAB~\cite{Agnese:2016cpb}, \textsc{DEAP-3600}~\cite{Adhikari:2020gxw}, \textsc{Damic-M}~\cite{Settimo:2018qcm}, \textsc{Darwin}~\cite{Aalbers:2016jon} and \textsc{Darkside-20k}~\cite{Aalseth:2017fik} serve as promising avenues not just for DM discovery, but, as we argue below, for reconstructing its astroparticle properties as well. 

\begin{figure}[!ht]
\centering
\includegraphics[width=0.3\textwidth]{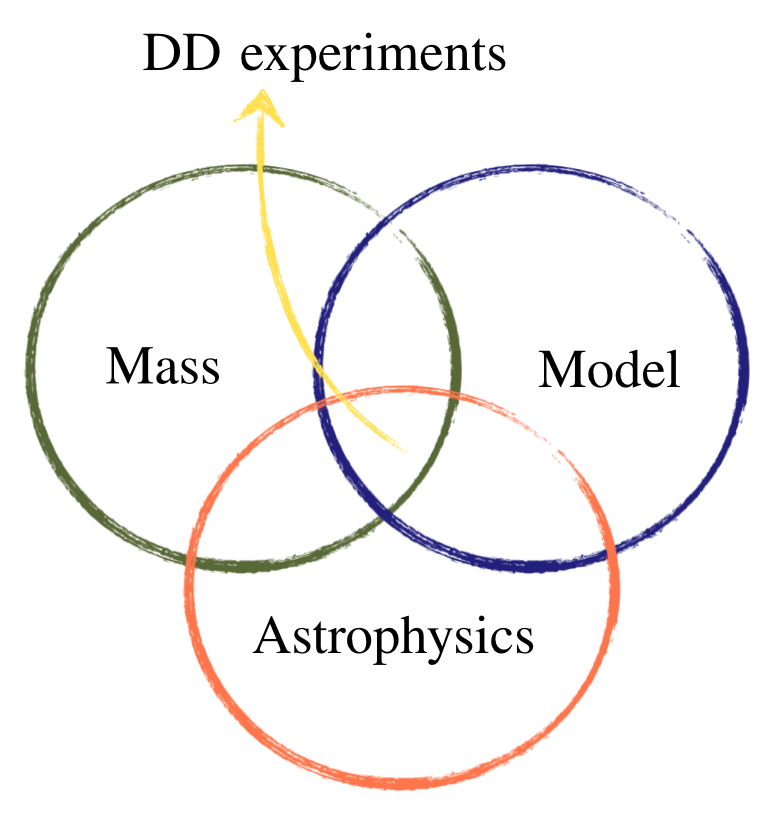}
\caption{Three-fold degeneracy in DM direct detection.}\label{fig:schematic_plot_1}
\end{figure}

The main physical observable in a DD experiment is the differential recoil spectrum, typically quoted as a function of the primary scintillation signal. Interestingly, modeling the DM recoil spectra at DD experiments relies on several independent aspects of its phenomenology. More specifically, DD experiments probe a combination of three important DM properties: its mass, interaction type (with nucleus and/or electrons) which we will refer to as the model, and its astrophysical distribution, namely through its density and velocity distribution in the solar neighborhood. From the perspective of statistical inference, this results in a three-fold degeneracy depicted in Fig.~\ref{fig:schematic_plot_1}. As corollary, determining the local astrophysical properties of DM precisely will be crucial in reconstructing its particle physics properties.

The Standard Halo Model (SHM) of DM velocity distribution~\cite{Drukier86} has been the cornerstone of DD analyses since it was proposed nearly three decades ago. The SHM follows from modeling the Milky Way (MW) as an isotropic, isothermal halo in equilibrium formed through the virialization of multiple subhalo merger residue. However, observations~\cite{1994Natur.370..194I, 1995ApJ...451..598J, 2001ApJ...551..294I, Belokurov:2006ms} and results from $N$-body simulations~\cite{Vogelsberger:2008qb, Kuhlen:2009vh, Kuhlen:2012fz, Kuhlen:2012ft, Mao:2013nda, Bozorgnia:2016ogo} have suggested the presence of diverse stellar and DM substructures from recent mergers, challenging the MW's steady state characterization. There have also been attempts to semi-analytically model the local DM velocity distribution using kinematic data~\cite{Lisanti:2010qx, Catena:2011kv, Bhattacharjee:2012xm, Bozorgnia:2013pua, Fornasa:2013iaa, Mandal:2018efq, Petac:2018wnu}, but these rely on additional, potentially restrictive, assumptions about the structure of the MW. 

On the other hand, astrometric data released by the European Space Agency's {\it Gaia} mission~\cite{Gaia2016, Gaia2018} presents a unique opportunity to study the MW's accretion history, and take first steps toward an empirical determination of the DM phase-space distribution. Even with a subset of the full data, a few groups~\cite{2018ApJ...863L..28M, Helmi2018, 2018ApJ...856L..26M, 2018ApJ...862L...1D} have reported evidence of tidal debris from a dominant merger in the solar neighborhood, the so-called {\it Gaia} Sausage or {\it Gaia} Enceladus, with very different kinematics compared to the old, virialized stellar population. While the full implication of stellar data for the astrophysical properties of DM will take decades to analyse, pioneering work by ref.~\citep{Necib:2018iwb} used the kinematics of MW halo stars in the cross-matched Sloan Digital Sky Survey (SDSS)-{\it Gaia} data set as tracers for the DM velocity, and validated their analysis~\cite{Necib:2018igl} with the \textsc{Fire-2} cosmological zoom-in simulation. Their analysis used a mixture model to characterize the local DM velocity distribution as a sum of the smooth halo and a radially aniostropic substructure components, which we refer to henceforth as the {\it Gaia} velocity distribution.

In this paper, we perform a systematic study of how the DM velocity distribution affects the reconstruction of DM model parameters at current or near-future DD experiments. Our analysis expands the existing DM direct detection literature in two important ways: 
\begin{itemize}
\item We consider the effect of DM velocity distribution for DM models that encompass a diverse set of operators in non-relativistic effective field theory (NREFT) with different recoil energy and velocity dependences. We show that the {\it Gaia} velocity distribution could significantly change not just the overall rate, but also the shape of the DM recoil spectrum at DD experiments.
\item Adopting the euclideanized signal (ES) method developed by refs.~\cite{Edwards:2017mnf, Edwards:2017kqw}, we forecast the ability of next-generation experiments to resolve DM model parameters using SHM and {\it Gaia} velocity distributions as two representative cases.\footnote{The ES method also conveniently aligns with the maxim,``methods flow from assumptions" --- one of David W. Hogg's {\it guiding} principles of data analysis. It is introduced around the 13:40 mark here: \url{https://physicslearning.colorado.edu/tasi/tasi_2018/videos/june11/June11Clip1.mp4}.}  We find that both the DM mass and the recoil energy dependence of the model could enhance or suppress the effect of the {\it Gaia} velocity distribution vis-\'{a}-vis the SHM while inferring the DM model parameters. This is another example of the curious interplay between different DM properties illustrated in Fig.~\ref{fig:schematic_plot_1}. 
\end{itemize}
Our approach is quite different from both traditional forecasting approaches for DM direct detection that used benchmark-dependent mock data sets~\cite{McDermott:2011hx, Catena:2014epa, Gluscevic:2014vga, Gluscevic:2015sqa, Gelmini:2018ogy}, or analyses which studied the effect of uncertainties in SHM on constraining the DM particle physics properties~\cite{Kahlhoefer:2017ddj, Wu:2019nhd, Besla:2019xbx, Hryczuk:2020trm}.

The paper is organized as follows. We review the basic ingredients to compute the recoil spectra at DD in section~\ref{sec:EFT}: we summarize and review the velocity distribution taking into account of the {\it Gaia} Sausage in section~\ref{sec:Astro-overview}, and review the NREFT formalism along with several benchmark DM models in section~\ref{sec:scatteringtheory}. In section~\ref{sec:DMvelocity}, we demonstrate how {\it Gaia} velocity distribution could modify both the overall rate and the shape of the recoil spectra in the NREFT framework. In Section~\ref{sec:method}, we introduce the recently developed ES statistical framework, which allows us to make forecasts without running MC simulations. In section~\ref{sec:results}, we present our results on the effects of {\it Gaia} distribution on reconstructing DM model parameters and distinguishing between different models at future DD experiments as well as setting constraints using current data. 
We conclude in section~\ref{sec:con}.

\section{Phenomenology of dark matter-nuclear Interaction}
\label{sec:EFT}
In this section, we briefly review the key ingredients to compute the rate of DM scattering off nuclei in a DD experiment. In particular, we summarize possible new DM velocity distributions inferred from the {\it Gaia} data. We also review both the model-independent framework, the non-relativistic effective theory and some specific benchmark models to study different types of DM-nucleus interactions. We emphasize that this section is a review of the literature, which are most relevant to our studies. Readers who are familiar with the subject could skip this section. A more extensive recent review on DD can be found in ref.~\cite{Schumann:2019eaa}.

\subsection{DM velocity distributions}
\label{sec:Astro-overview}
Our key experimental observable, the differential recoil rate (the full formula is provided in Appendix~\ref{sec:dd_basics}), is sensitive to the DM velocity distribution. In this section, we first review the SHM and the {\it Gaia} distribution which we use in our analysis and then compare the velocity moments from different possible velocity distributions.  

The local velocity distribution in the GC frame for the SHM, {\it i.e.} for an isotropic, isothermal DM halo in equilibrium, is well-modeled by a truncated Maxwell-Boltzmann distribution~\cite{Drukier86}, 
\beq\label{eq:SHM}
f_{\rm SHM}(\vec{v}) =  \frac{1}{N_{\rm esc}}\frac{1}{(2\pi\sigma_v^2)^{3/2}}e^{-v^2/2\sigma_v^2} \, \Theta(v_{\rm esc} - v),
\eeq
where we take the velocity dispersion and escape velocity of the DM halo to be $\sigma_v \approx 160$ km/s and $v_{\rm esc} \approx 540$ km/s respectively~\cite{Aprile:2017iyp}, and the normalization constant $N_{\rm esc}$ is given by,
\beq\label{eq:norm}
N_{\rm esc} = {\rm erf}\left[ \frac{v_{\rm esc}}{\sqrt{2} \sigma_v} \right] - \frac{2}{\sqrt{\pi}} \, \left( \frac{v_{\rm esc}}{\sqrt{2} \sigma_v} \right) e^{-v_{\rm esc}^2/2\sigma_v^2} \, .
\eeq
The DM velocity distrubution in the Earth frame, $\tilde{f}(\vec{v})$, can be obtained by boosting GC frame distribution, $f(\vec{v})$, with the Earth's velocity, ${\vec v}_{\rm obs} (t)$, 
\beq
\tilde{f}(\vec{v}) = f\left(\vec{v}_{\rm obs} (t) + \vec{v}\right). 
\eeq
where the $t$ dependence arises due to Earth's orbital motion around the Sun and is commonly referred to as annual modulation~\cite{Savage:2006qr, Lee:2013xxa}. Ignoring modulation effects, we assume a time averaged value for the Earth's velocity ${v_{\rm obs} \approx 230 \, {\rm km/s}}$ in our analysis.

A crucial assumption in the formulation of the SHM is the condition of local equilibrium. However, if the MW has undergone one or more recent mergers, the equilibrium condition is then invalid, and we would need a method to empirically determine the DM velocity distribution. Assuming the CDM paradigm, hierarchical structure formation implies that the MW halo should primarily consist of virialized tidal debris from old subhalo mergers with other spatial and kinematic substructure sourced by more recent ones. Ref.~\cite{Herzog-Arbeitman:2017fte} used the \texttt{Eris} $N$-body simulation to show the correlation between velocities of old, metal-poor stars and the virialized DM component of a MW-like halo. Along the same lines, ref.~\cite{Kuhlen:2012fz} argued that the velocity distribution of stars in a class of substructure called {\it debris flow}\footnote{Debris flow consists of tidal debris of an accreted dwarf galaxy that has made several orbits such that it is spatially mixed on large scales while retaining a unique signature in velocity space~\cite{Lisanti:2011as}.} is a good kinematic tracer of its accreted DM counterpart based on the \texttt{Via Lactea} simulation. 

In unarguably the golden-age of data-driven astrophysics, we can now obtain 7D information\footnote{Including parallax, sky positions and proper motions on the celestial sphere, and radial velocity measurements from {\it Gaia}, along with metallicity data from SDSS. The stellar metallicity, given by the iron-to-hydrogen abundance ratio, $[{\rm Fe}/{\rm H}]$, is used as a proxy for the star age.} for main-sequence stars in the MW halo by cross-matching the \Gaia data releases~\cite{Gaia2016, Gaia2018} with the Sloan Digital Sky Survey (SDSS). Using this cross-matched catalog, refs.~\cite{2018ApJ...863L..28M, Helmi2018} found signatures of a debris flow in the solar neighborhood (within $\sim 4$ kpc of the Sun)-- the so-called {\it Gaia}-Sausage -- that consists of metal-rich halo stars with high radial anisotropy. Further investigation of other phase space substructures~\cite{2018ApJ...856L..26M, 10.1093/mnras/sty1403, 2018ApJ...863L..28M} indicates that, in fact, the MW might have experienced at least two different accretion events, namely those of the {\it Sausage}~\cite{2018ApJ...863L..28M} and {\it Sequoia}~\cite{2019ApJ...870L..24B} dwarf galaxies. The {\it Gaia}-Enceladus structure~\cite{Helmi2018} hints at a possible third event, although it appears to partially consist of debris from the other two mergers~\cite{2019MNRAS.488.1235M}.

\begin{figure}[!h] 
    \centering
      \begin{subfigure}[!h]{0.45\textwidth}
        \includegraphics[width=\textwidth]{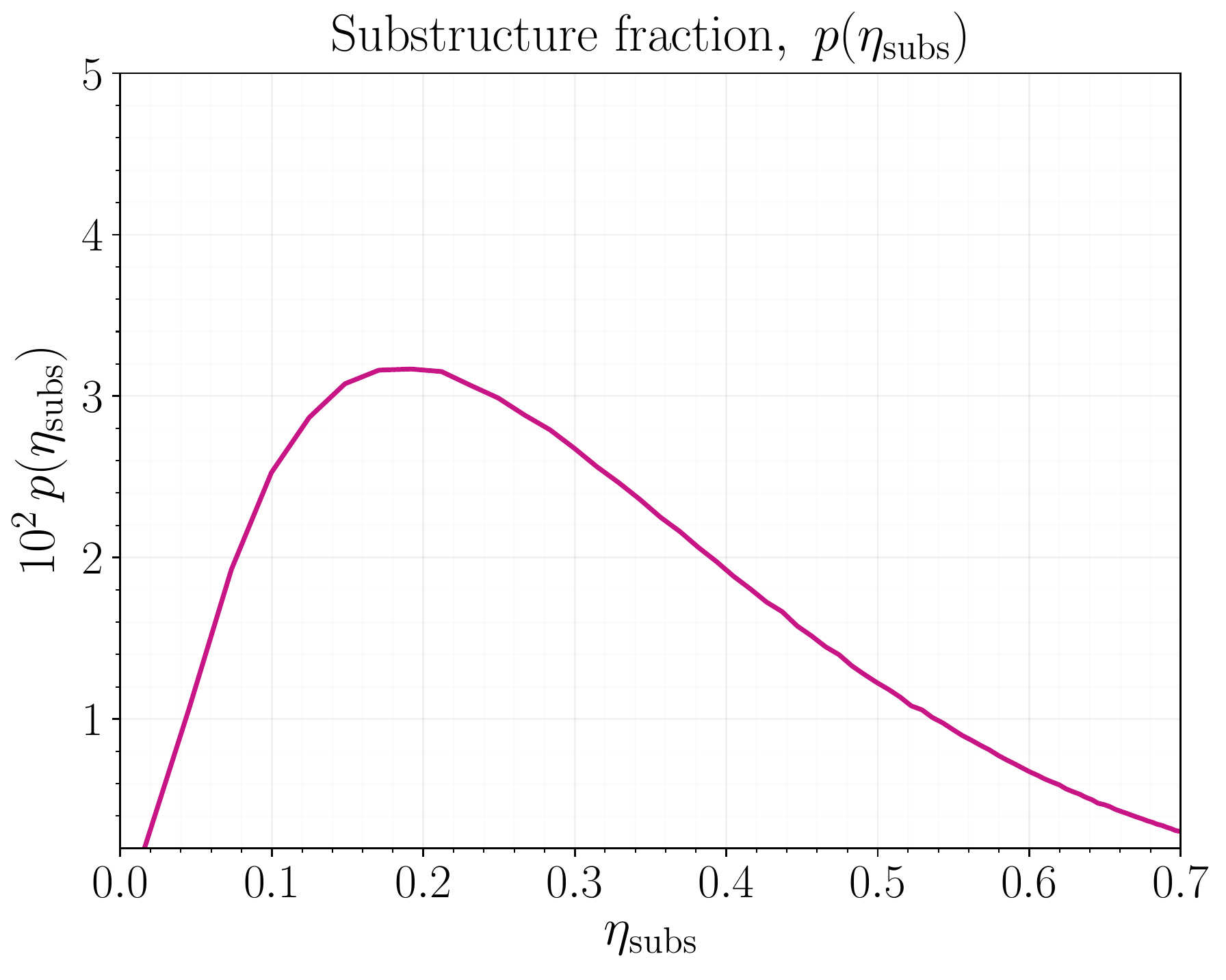}
    \end{subfigure}
    ~ \quad
    \begin{subfigure}[!h]{0.45\textwidth}
        \includegraphics[width=\textwidth]{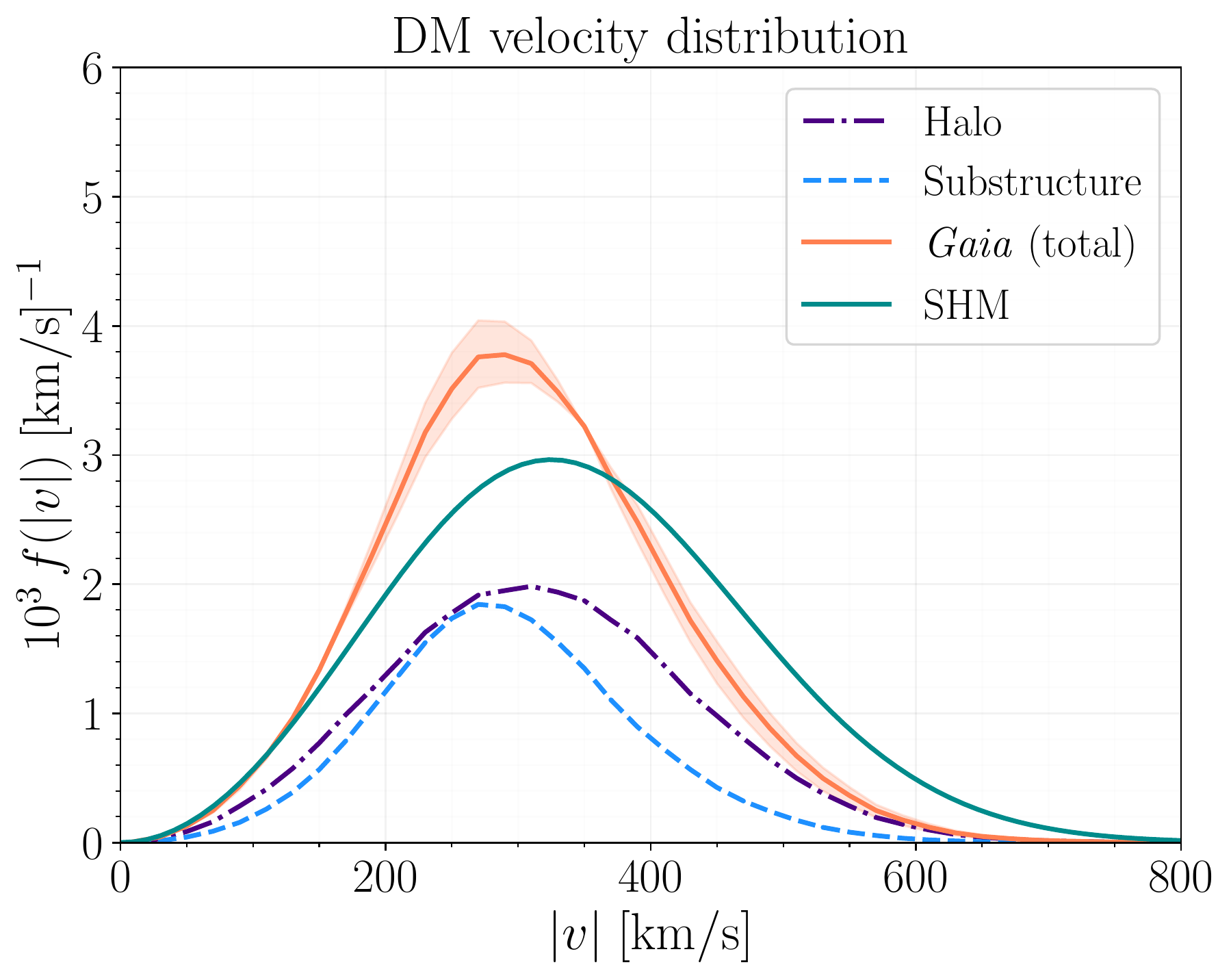}
    \end{subfigure}
\caption{Left: distribution of the DM fraction in substructure $\eta_{\rm sub}$ determined using an empirical relation between the mass-to-light ratio and metallicities for the MW (provided by L. Necib in private communication). Right: the normalized {\it Gaia} DM velocity distributions for the halo (indigo dash dotted) and substructure (blue dashed) components weighted by the median DM substructure fraction, {\it i.e.} $\eta_{\rm sub} = 0.42$. Also shown for reference are the total velocity distributions in the heliocentric frame for {\it Gaia} for the median value of DM fraction (orange solid) and the SHM (cyan solid). The band of the {\it Gaia} distribution is obtained by varying $\eta_{\rm sub}$ in the $1\sigma$ range around its median value.} \label{fig:DM_f_v}
\end{figure}

For our analysis, we only focus on the effects of the {\it Gaia}-Sausage, since it is the dominant merger in the solar neighborhood contributing $\sim 70$\% of all accreted low-metallicity stars~\cite{2018ApJ...863L..28M}. Ref.~\cite{Necib:2018iwb} used the SDSS-{\it Gaia} DR2 data set with a subsample of $\sim 190,000$ stars to make the first empirical determination of the local DM velocity distribution. They performed a Gaussian mixture model (GMM) analysis on the joint distribution of stellar velocities and metallicities to classify stars in three populations with distinct kinematic properties: metal-rich young disk stars formed {\it in-situ}; accreted stars which include metal-poor stars in the smooth isotropic halo, and intermediate metallicity stars with a high radial anisotropy that constitute the {\it Gaia}-Sausage substructure. The total DM velocity can then be written as a linear combination of the substructure and halo velocities, weighted by the the fraction of DM in each component,
\beq\label{eq:sub_frac}
f(v) =  (1 - \eta_{\rm sub}) f_{\rm halo}(v) + \eta_{\rm sub} f_{\rm sub}(v), 
\eeq
where $\eta_{\rm sub}$ parametrizes the DM fraction in substructure. By sampling the GMM model with the SDSS-{\it Gaia} DR2 subsample, ref.~\cite{Necib:2018iwb} inferred the best-fit velocity distribution for each component\footnote{Available publicly here: \url{https://github.com/linoush/DM_Velocity_Distribution}.} as well as a posterior distribution of their stellar fractions. Subsequently, ref.~\cite{Necib:2018igl} derived an approximate relation between the mass-to-light ratio and metallicity for the MW~\cite{2013ApJ...779..102K, Garrison-Kimmel:2016szj} for using the stellar fraction posterior to estimate the DM fraction distribution, $p(\eta_{\rm sub})$, shown in the left panel of Fig.~\ref{fig:DM_f_v}.\footnote{While ref.~\cite{Necib:2018igl} only provides the median value of $\eta_{\rm sub}$ with its $1\sigma$ error bar, the full distribution was provided to us by one of the authors, L. Necib, in private communication.} Importantly, however, the fraction of stars in substructure relative to total accreted stars is typically a poor tracer for $\eta_{\rm sub}$ as simulations show that the halo DM component may contain significant contributions from accretion of diffuse DM and DM in non-luminous subhalos. In fact, as demonstrated by ref.~\cite{Necib:2018igl, Bozorgnia:2018pfa, Bozorgnia:2019mjk}, late-time accretion from the latter component may affect both the shape and the velocity distribution of all accreted DM. We postpone a detailed study of these effects to future work.

We plot the {\it Gaia} DM velocity distribution in the heliocentric frame for each component from ref.~\cite{Necib:2018iwb} in the right panel of Fig.~\ref{fig:DM_f_v}. We also note that there is no smooth interpolation in $\eta_{\rm sub}$ between the SHM and the {\it Gaia} velocity distribution, {\it i.e}. $\eta_{\rm sub} = 0$ does not yield the SHM.\footnote{We note here that ref.~\cite{Evans:2018bqy} uses the sphericity constraint of the DM halo~\cite{2019MNRAS.485.3296W} to argue that $\eta_{\rm sub} \lesssim 20$\%. However, this constraint relies on the assumption that the observed stellar density of the Sausage is a reasonably good tracer for the density of its DM component. In any case, as we show in fig.~\ref{fig:sum_plot}, the qualitative behavior of our result holds for a wide range of substructure fractions.} Heuristically, the differences between the two velocity distributions can be attributed to the {\it Gaia} one being inferred through a better statistical modeling of the same stellar population. 

There has also been growing interest in studying the effect of other phase-space substructures in the solar neighborhood discovered using {\it Gaia} data, in particular for the retrograde S1~\cite{10.1093/mnras/sty1403}, and the prograde Nyx~\citep{Necib:2019zka, Necib:2019zbk} streams. Stellar streams appear as a coherently moving group of stars resulting from the tidal debris of a galaxy localized in both position and velocity space, and presence of a significant DM fraction in stream(s) could result in a very different annual modulation signature compared to the SHM. With this motivation, ref.~\cite{Buckley:2019skk} revisited the DM interpretation of the latest DAMA data with the S1 stream, and found that absence of a DM signal at other experiments rules out the preferred DM parameter space of DAMA even if $100\%$ of the local DM was present in such a stream. Another promising avenue to look for interesting signatures of DM substructure are axion searches and directional detection experiments as discussed in refs.~\cite{Evans:2018bqy, OHare:2018trr, OHare:2019qxc}. The main drawback of the aforementioned analyses is the underlying assumption of a near perfect stellar-DM velocity correlation. As illustrated in ref.~\cite{Necib:2018igl} (see top panel of Fig. 7 for example), stellar streams turn out to be poor tracers of the DM velocity, since the tidal debris in the stream hasn't had enough time to completely mix with the halo. In addition, compared to the Sausage substructure, streams are expected to contribute only a subdominant fraction of DM in the solar neighborhood. For example, using the results of ref.~\cite{2017MNRAS.464.2882A}, ref.~\cite{OHare:2019qxc} argued that the progenitors of the S1 and S2 stream could contribute ${\sim} 1$-$10$\% of the local DM fraction. Thus, we only focus on the DM velocity distribution associated with the {\it Gaia} Sausage based on refs~\cite{Necib:2018iwb, Necib:2018igl} in our work, and ignore the effect of streams in our analysis. 

\begin{figure}[h]
    \centering
        \begin{subfigure}[!h]{0.44\textwidth}
        \includegraphics[width=\textwidth]{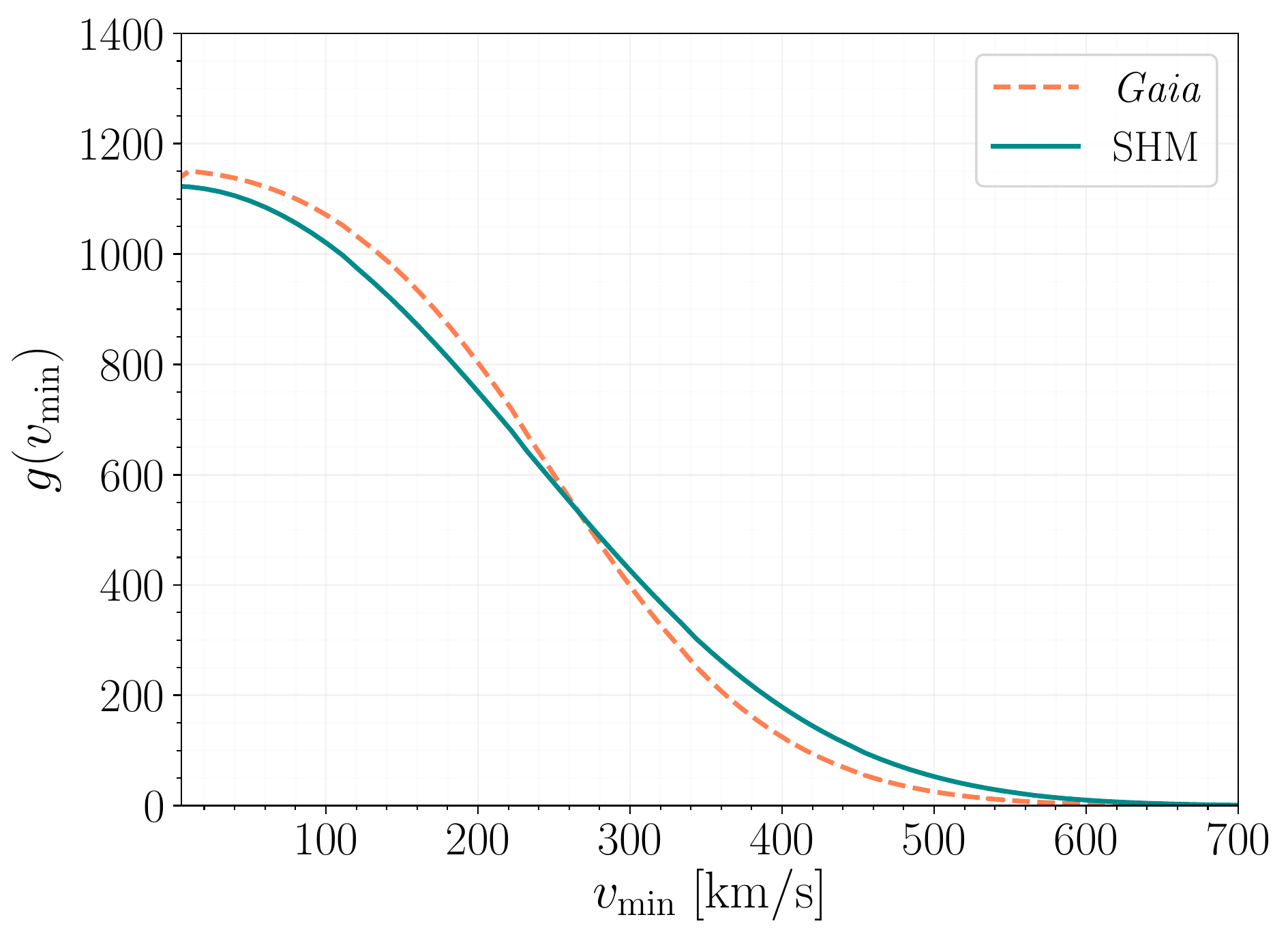}
        \caption{$g(v_{\rm min})$}
        \label{subfig:g_vmin}
    \end{subfigure}
    ~ \qquad
    \begin{subfigure}[!h]{0.44\textwidth}
        \includegraphics[width=\textwidth]{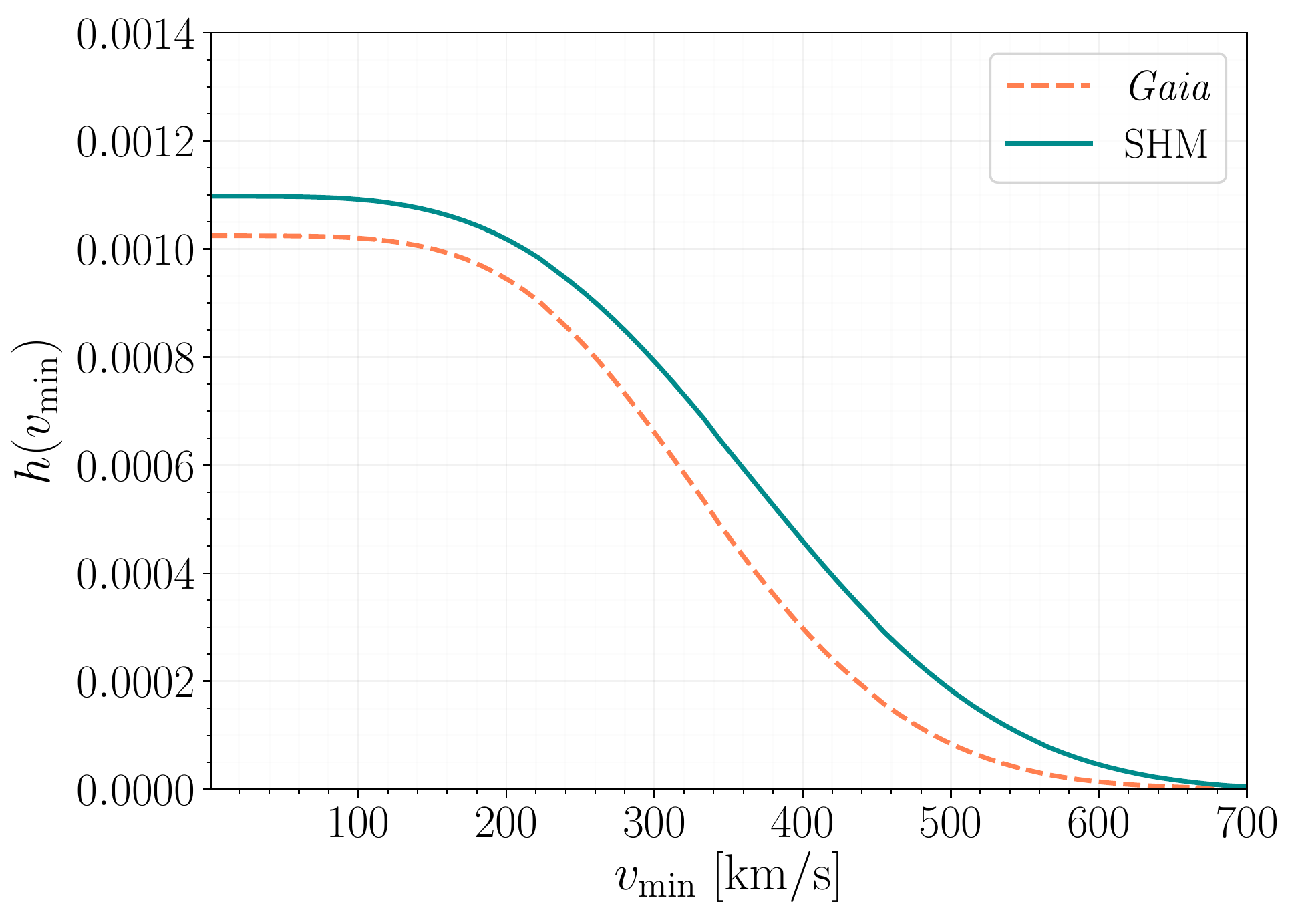}
        \caption{$h(v_{\rm min}$)}
        \label{subfig:h_vmin}
    \end{subfigure}
        ~ \qquad
    \caption{Moments of the empirical {\it Gaia} velocity distribution (orange, dashed) and the SHM distribution (cyan, solid) relevant in our analysis, $g(v_{\min})$ (left)  and $h(v_{\min})$ (right), as functions of $v_{\min}$. The value of $g(v_{\min})$ ($h(v_{\min})$) has been multiplied (divided) by the speed of light to be made dimensionless, for a better illustration of their relative magnitudes.}\label{fig:f_v_int}
\end{figure}

Dark matter interactions with the nucleus beyond the simplest contact one could introduce additional factors of incident DM velocity squared $v^2$ into the velocity integration. We will consider two velocity moments: 
\beqa\label{eq:g_v}
g(v_{\rm min}) & \displaystyle = \int^{v_{\rm esc}}_{v_{\min}}\dd^3v \frac{\tilde{f}(\vec{v})}{v}, \\
h(v_{\rm min}) & \displaystyle = \int^{v_{\rm esc}}_{v_{\min}}\dd^3v \,v \tilde{f}(\vec{v}), \label{eq:h_v}
\eeqa
where $v_{\rm min} = \sqrt{\frac{m_T E_R}{2 \mu_T^2}}$ is the minimum DM velocity for a given recoil energy $E_R$, target mass $m_T$, DM-nucleus reduced mass $\mu_T$. The velocity integrals are bounded from above by the escape speed, $v_{\rm esc}$, which we take to be $\infty$ for the {\it Gaia} velocity distribution following ref.~\cite{Necib:2018iwb}. 

We plot $g(v_{\rm min})$ and $h(v_{\rm min})$ as functions of $v_{\rm min}$ in Fig.~\ref{fig:f_v_int}. From the left panel of Fig.~\ref{fig:DM_f_v}, we note that $f_{\it Gaia}$ peaks at a lower $v$ and has a smaller high velocity tail compared to $f_{\rm SHM}$. Integrating it with $1/v$ results in $g(v_{\rm min})_{\it Gaia} > g(v_{\rm min})_{\rm SHM}$ for small $v_{\rm min}$, whereas for $v_{\rm min} \gtrsim 250$ km/s, $g(v_{\rm min})_{\it Gaia} < g(v_{\rm min})_{\rm SHM}$. Meanwhile in case of $h(v_{\rm min})$, the preference for large velocities due to an additional factor of $v$ results in $h(v_{\rm min})_{\it Gaia} < h(v_{\rm min})_{\rm SHM}$ for all values of $v_{\min}$. These features are crucial for us to understand how the {\it Gaia} distribution affects the recoil spectra at the qualitative level in Sec.~\ref{sec:DMvelocity}. 

\subsection{DM-nucleus scattering theory}
\label{sec:scatteringtheory}

In this section, we will briefly review a model-independent framework and some concrete models to study different types of DM scattering off nucleus in DD experiments. There is a huge literature on possible DM scattering in DD and we do not intend to provide an exhaustive review here. We will only refer the reader to the original papers and papers we actually use for our analysis.


The typical DM velocity in our galaxy is $v \lsim 10^{-3}c$. The incident DM kinetic energy is around $\mathcal{O}(10)$ keV for a DM particle with mass $\sim 10$ GeV. Given that DM scattering in DD experiments is non-relativistic (NR), a simple model-independent way to parametrize different types of DM-nucleon interaction is the Galilean-invariant NR effective field theory (NREFT), first proposed and developed in refs.~\citep{Fan:2010gt} and~\citep{Fitzpatrick:2012ix}. The core result is that in the NR limit, DM-nucleon interactions could be encoded in 16 NREFT operators, 15 of which are linearly independent. These operators are expressed in terms of four three-vectors: DM spin $\vec{S}_\chi$, nuclear spin $\vec{S}_N$, the momentum transfer $\vec{q}=\vec{p}^{\,\prime}-\vec{p}$ with $\vec{p}$ $(\vec{p}^{\,\prime})$ the incoming (outgoing) DM three-momentum and the transverse velocity
\beq
\vec{v}^\perp = \vec{v} + \frac{\vec{q}}{2\mu_T}, \quad {\rm where~}\; \vec{v}^\perp \cdot \vec{q} = 0.
\eeq
We will use only 12 linearly independent operators, $\OO_i$'s, listed in Table~\ref{table:Operators}. These 12 operators are usually sufficient to describe the NR limit of many relativistic operators that appear in simple models with spin-0 or spin-1 mediators. In this paper, we only focus on spin-independent scattering.

\begin{table}[!htb]
\centering
\begin{tabular}{| c | c | c |  }
\hline Operators & Form & Spin-Dependence  \\ \hline
$\OO_{1}$ & $\mathbbm{1}$ & \xmark \\
$\OO_{2}$  & $(\vec{v}^{\perp})^2$ & \xmark   \\
$\OO_{3}$ & $ i \vec{S}_    N \cdot (\vec{q}\times\vec{v}^\perp) $ & \checkmark \\
$\OO_{4}$& $ \vec{S}_\chi\cdot\vec{S}_N $ &  \checkmark   \\
$\OO_{5}$ & $i\vec{S}_\chi \cdot (\vec{q}\times\vec{v}^\perp)$ & \xmark  \\
$\OO_{6}$ & $(\vec{S}_N\cdot\vec{q})(\vec{S}_\chi\cdot\vec{q})$ & \checkmark \\
$\OO_{7}$ & $\vec{S}_N\cdot\vec{v}^\perp$ & \checkmark \\
$\OO_{8}$ & $\vec{S}_\chi\cdot\vec{v}^\perp$ & \xmark   \\
$\OO_{9}$ & $i\vec{S}_\chi \cdot (\vec{S}_N\times\vec{q})$  & \checkmark  \\
$\OO_{10}$ & $i\vec{S}_N\cdot \vec{q}$ & \checkmark  \\
$\OO_{11}$ & $i\vec{S}_\chi\cdot \vec{q}$ & \xmark  \\ 
$\OO_{12}$ & $\vec{v}^\perp \cdot (\vec{S}_\chi \times \vec{S}_N)$ & \xmark  \\ \hline
\end{tabular}
\caption{Summary of the NREFT operators. The second column indicates the operators' nuclear spin dependence. }
\label{table:Operators}
\end{table}

In the NR limit, a relativistic operator in the field theory can be mapped onto a linear combination of NR operators. Thus, a relativistic Lagrangian for a particular DM scattering model, which could contain several relativistic operators for DM-nucleon interaction, can be written in terms of the NREFT operators as,
\beq\label{eq:L_map}
\mathcal{L}_{\rm NREFT} = \sum_{N=n,p} \sum_{i=1}^{12}\frac{c^{(N)}_i\OO^{(N)}_i}{q^2 + m_{{\rm med};i}^2}, 
\eeq
where $N = n, p$ labels the type of nucleon DM interacts with, which could be either neutron or proton and $i$ labels NREFT operators. The coefficients, $c_i$'s, depend on the coupling coefficients in the relativistic theory and the Wilson coefficients obtained by mapping the relativistic operators to the NR ones. Compared to the standard literature (e.g., refs~\cite{Fitzpatrick:2012ix, DelNobile:2013sia, DelNobile:2018dfg}), we also take the mediator propagator out of $c_i$s and explicitly write it out, where $m_{{\rm med};i}$ is the mediator mass for the $i$th interaction. Strictly speaking, the formula above holds when the mediator is light with mass below GeV. When the mediator is heavy with mass above GeV, it could be integrated out and the propagator in eq.~\eqref{eq:L_map} is reduced to $m_{\rm med}^{-2}$.

In the paper, we only consider the leading order spin-independent (SI) elastic scattering in which DM scatters off the entire nucleus coherently. To go from DM scattering with individual nucleons to scattering with nucleus, one needs to take into account of the nuclear response which is encoded in the form factor 
\beq
\FF_{i,j}^{(N,N')} &=& \big\langle \,\text{Nucleus}\, \big|\, \OO_i^{(N)} \OO_j^{(N')} \big| \,\text{Nucleus}\, \big\rangle, \nonumber \\
&=& \sum_{n=0}^\infty F^{(n)}_{i,j}(q^2) v^{2n},
\eeq
where in the second line, we expand it as a power series of $v^2$ and $F^{(n)}$ are associated coefficients. Further development of effective field theory from quarks to nucleons could be found in refs~\cite{Bishara:2016hek, Bishara:2017pfq}. 
In NREFT, the differential scattering rate (full formula in Eq.~\eqref{eq:rate_short}) in terms of the $c_i$s and the form factors is,
\beq\label{eq:rate}
\frac{\dd R}{{\dd} E_R} &\propto & \sum_{\substack{i,j, \\N,N'}}\int^{\infty}_{v_{\min}} \dd^3 v\frac{\tilde{f}(\vec{v})}{v}\frac{c^{(N)}_i  c^{(N')}_j}{(q^2 + m_{{\rm med};i}^2)(q^2 + m_{{\rm med};j}^2)}  \FF^{(N,N')}_{i,j}(q^2, v^2), \nonumber \\
&=& \sum_{i,j} \frac{c_i  c_j}{\left(q^2+m_{{\rm med};i}^2\right)\left(q^2+m_{{\rm med};j}^2\right)}  \left[ g(v_{\min}) F^{(0)}_{i,j}(q^2) + h(v_{\min}) F^{(1)}_{i,j}(q^2) + \cdots \right].
\eeq
Note that our normalization of the form factors $\FF$'s differs from that in the literature (e.g., refs~\cite{Fitzpatrick:2012ix, DelNobile:2013sia}) by a factor of $(4 m_\chi m_N)^2$ with $m_N$ the nucleon mass. The form factors used in our analysis are provided in Appendix~\ref{sec:NREFT_operator}.

A DM model could contain several different DM-nucleon interactions and a relativistic operator between DM and nucleons could map onto multiple NR operators. Moreover, the constraint on the coupling of the relativistic operator could be a complicated sum of constraints on each NR operator it corresponds to with different weights due to interferences between different NR operators. Thus in analyzing how the new velocity distribution affects interpretation of DD data, we also consider a few simple representative models. Again, we do not intend to be exhaustive; we only select and compare a few models with different $E_R$ and velocity dependences, which could be affected by SHM and {\it Gaia} velocity distributions in different ways. 

We assume for simplicity that the DM is fermionic, and analyze six SI DM models: DM interacting through heavy scalar mediator (leading to the simplest contact interaction), millicharged DM (mC), DM with magnetic dipole moment (MD) with either heavy or light mediators, DM with electric dipole moment (ED) with light mediator and anapole DM with a heavy mediator. All the models are summarized in table~\ref{table:SI_models}. More details of the models could be found in Appendix~\ref{sec:models}. 

\begin{table}[!h]
\centering
\def\arraystretch{2.0}
\begin{tabular}{| c | c | c |  c |  }
\hline Model & Relativistic Operator & NREFT Operator & $E_R$ and DM velocity moment \\ \hline
\pbox{5cm}{ Contact interaction \\ (heavy gauge boson mediator)} & \pbox{10cm}{$g_c\bar{\chi}\gamma_\mu\chi\bar{N}\gamma^\mu N$} & $\displaystyle  g_{c} \OO^{(N)}_1 $ & $\sim g(v_{\rm min})$ \\[3pt]
\pbox{5cm}{Millicharge (mC) \\ w/ light mediator} & $\displaystyle  e\epsilon_\chi \bar{\chi}\gamma^\mu\chi A_\mu $ & $\displaystyle  e^2 \epsilon_\chi \frac{1}{q^2}\OO^{(p)}_1$ & $\sim E_R^{-2} \, g(v_{\rm min})$ \\[3pt]
\pbox{5cm}{Magnetic dipole (MD) \\ w/ heavy mediator} & $\displaystyle  \frac{\mu_\chi}{2} \bar{\chi} \sigma^{\mu\nu} \partial_\mu\chi \partial^\alpha F_{\alpha\nu}$ & $\displaystyle  \frac{e\mu_\chi}{2} \big( \frac{q^2}{m_\chi}\OO^{(p)}_1 - 4\OO^{(p)}_5\big)$ &  $\sim E_R^{2} \, g(v_{\rm min}) + E_R \, h(v_{\rm min}) $ \\[3pt]
\pbox{5cm}{Magnetic dipole (MD) \\ w/ light mediator} & $\displaystyle  \frac{\mu_\chi}{2} \bar{\chi} \sigma^{\mu\nu}\chi F_{\mu\nu} $ & $\displaystyle  \frac{e\mu_\chi}{2} \big( \frac{1}{m_\chi}\OO^{(p)}_1 - \frac{4}{q^2}\OO^{(p)}_5 \big) $ & $ \sim g(v_{\rm min}) +E_R^{-1} h(v_{\rm min}) $ \\[3pt]
\pbox{5cm}{Electric dipole (ED) \\ w/ light mediator}  & $\displaystyle  i\frac{d_\chi}{2}\bar{\chi}\sigma^{\mu\nu}\gamma^5 \chi F_{\mu\nu}  $ &  $\displaystyle  2ed_\chi \frac{1}{q^2}\ \OO^{(p)}_{11} $ &  $\sim E_R^{-1} \, g(v_{\rm min})$ \\[3pt]
\pbox{5cm}{Anapole \\ w/ heavy mediator} & $\displaystyle  i g_{\rm ana}\bar{\chi}\gamma^\mu\gamma_5 \chi \partial^\nu F_{\mu\nu}$ & $\displaystyle 2 eg_{\rm ana} \OO_8^{(p)} $& $\sim E_R \, g(v_{\rm min}) + h(v_{\rm min})$ \\[3pt]
\hline
\end{tabular}
\caption{Summary of the representative DM models used in our analysis. For brevity, only the SI NR operators of each model are shown here. We also show the leading order $E_R$ and velocity moment dependences of the corresponding spectra. The coupling constant in front of the operator defines the model parameter to be constrained, e.g. coupling strength $g_c$ for contact interaction, charge fraction $\epsilon_\chi$ for millicharged DM, and the dipole moment $\mu_\chi$ for magnetic dipole DM.}
\label{table:SI_models}
\end{table}

\section{Effect of the DM velocity distribution on DD recoil spectrum }
\label{sec:DMvelocity}
Before presenting the statistical method for DD forecasting and the results, we want to discuss how different DM velocity distributions could affect the differential recoil spectrum, the key quantity in DD experiments. Here we present results using NREFT operators. This will help us obtain some qualitative ideas and physical intuition of the effects of DM velocity distribution on interpreting current and future DD data in the context of a full-fledged model, which we present in section~\ref{sec:results}.

\begin{figure}[h]
\centering
\includegraphics[width=0.99\textwidth]{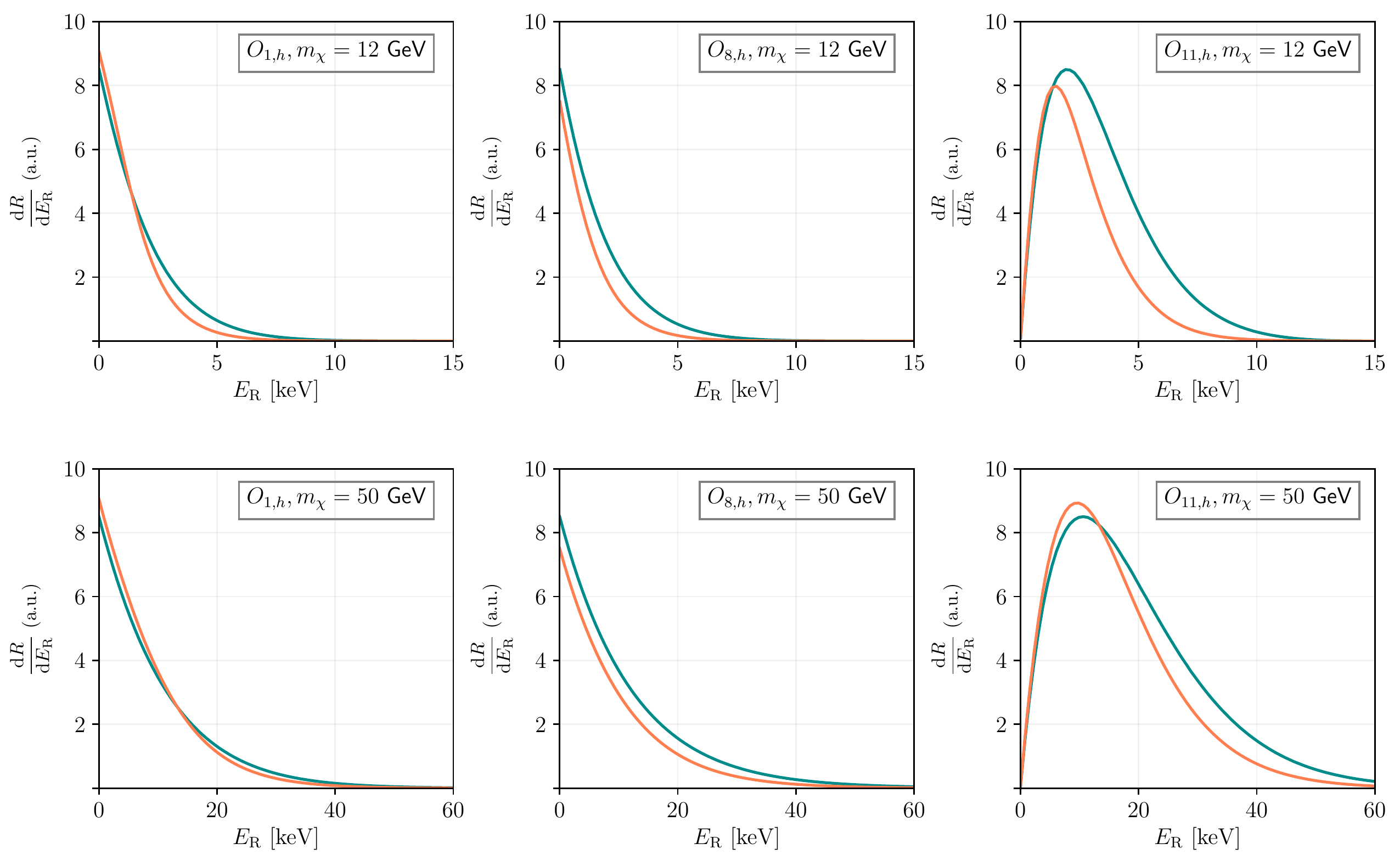} 
\caption{Top row: recoil spectra of $\OO_1$, $\OO_8$ and $\OO_{11}$ with heavy mediators for DM with $m_\chi = 12$ GeV. Bottom row: recoil spectra of the same set of operators for $m_\chi = 50$ GeV. The cyan (orange) curve assumes SHM (\Gaia) velocity distributions. In this figure and the next one, the differential recoiling rate is in arbitrary unit.}\label{fig:recoil_spectra_heavy}
\end{figure}

\begin{figure}[h]
\centering
\includegraphics[width=0.99\textwidth]{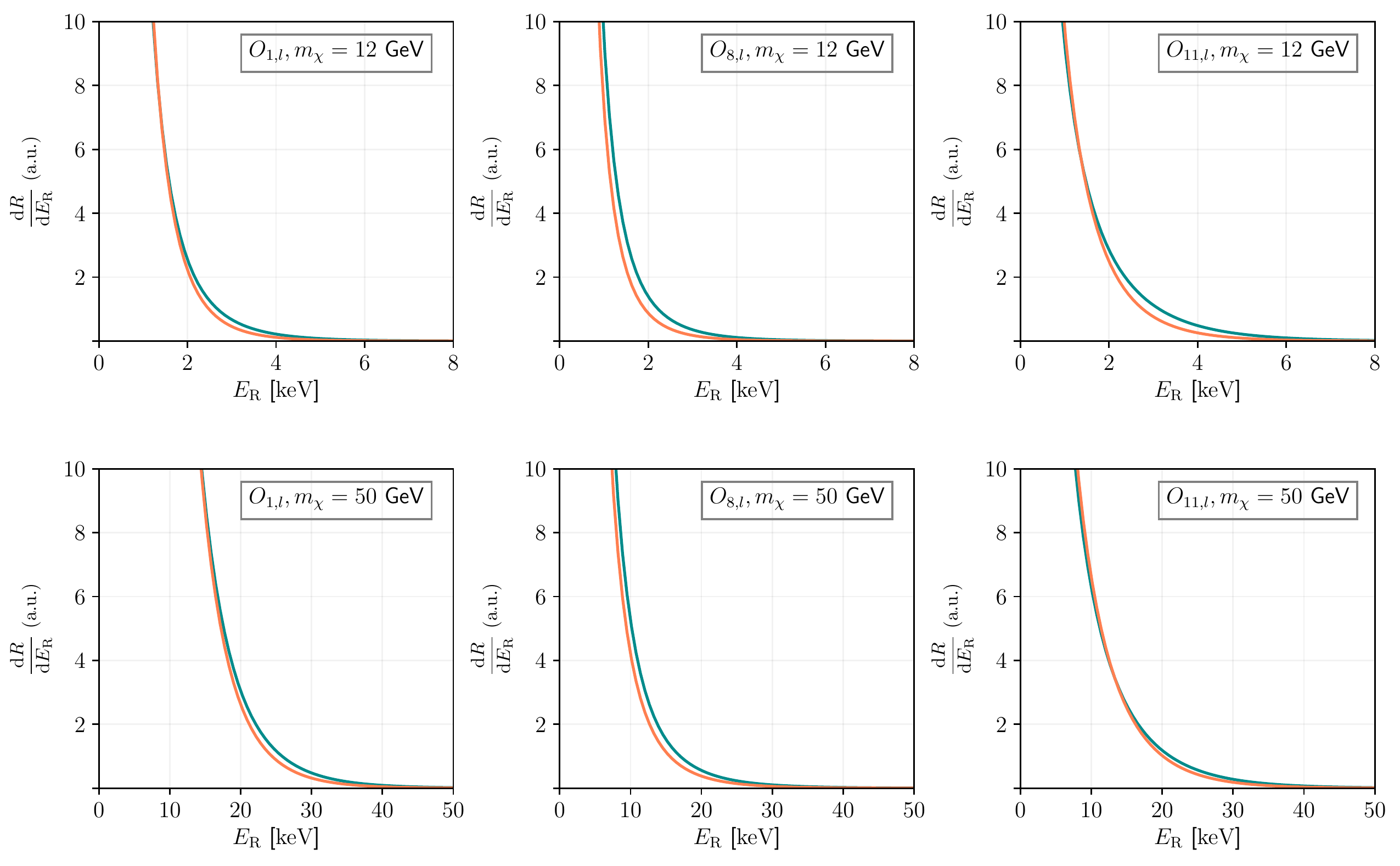}
\caption{Recoil spectra of $\OO_1$, $\OO_8$ and $\OO_{11}$ for DM with mass $m_\chi = 12$ (top) and 50 (bottom) GeV. Here we consider operators with light mediators. The cyan (orange) curve assumes SHM (\Gaia) velocity distributions.}\label{fig:recoil_spectra_light}
\end{figure}

The recoil spectrum depends on not only the velocity moments but also the coefficients $F^{(n)}_{i,j}(q^2)/(q^2+m_{\rm med}^2)^2$. The form factor function $F^{(n)}_{i,j}(q^2)$ is, in general, a polynomial of $q^2$ multiplying an exponent suppression factor $\sim e^{-q^2s^2}$, in which the effective nuclear radius is $s\approx 0.9$ fm. Since the exponential factor is common for all nuclei, we will focus on the polynomial part of $F^{(n)}_{i,j}(q^2)$ that depends on the type of DM-nucleon interaction.

The momentum transfer is related to the recoiling energy $E_R$ and $v_{\min}$ as, 
\beq
q^2 = 2m_T E_R= 4\mu_T^2 v_{\min}^2. 
\label{eq:qv}
\eeq
We could express $q^2$ in terms of $v_{\min}$ and the differential recoil rate as a function of $v_{\min}$ only. To illustrate how the recoil spectra vary with velocity distributions, we select three NR operators as examples
\beq\label{eq:op_example_response}
\OO_1 = \mathbbm{1}, \qquad \OO_8 = \vec{S}_\chi\cdot\vec{v}^\perp, \qquad \OO_{11} = i\vec{S}_\chi\cdot\vec{q},
\eeq
and compute their recoil spectra assuming a xenon target. For simplicity, we consider only one NR operator at a time with a single type of mediator. Let's consider two limits to illustrate the effects of different DM velocity distributions: heavy mediator with $m_{\rm med}^2 \gg q^2$ so that the propagator is approximately a constant, and light mediator with $m_{\rm med}^2 \ll q^2$ so that the propagator is approximately $1/q^2$. It is not difficult to generalize the discussion to cases in between the two limits with $m_{\rm med}^2 \sim q^2$. In the two limits, we have the dependence of the recoil spectrum on $v_{\min}$ for each operator schematically as
\beqa
\OO_1&:& g(v_{\min}), \qquad   \OO_8: v_{\min}^2 g(v_{\min}) + h(v_{\min}), \qquad  \OO_{11}:  v_{\min}^2 g(v_{\min}), \quad {\rm heavy \, mediator};  \nonumber \\
\OO_1&:& \frac{g(v_{\min})}{v_{\min}^4}, \qquad   \OO_8: \frac{g(v_{\min})}{v_{\min}^2} + \frac{h(v_{\min})}{v_{\min}^4}, \qquad  \quad \; \OO_{11}:  \frac{g(v_{\min})}{v_{\min}^2}, \qquad \quad {\rm light \, mediator}.
\eeqa
To derive the spectrum's dependence on $v_{\min}$ for $\OO_8$, we use $\vec{v}^\perp = \vec{v} + \frac{\vec{q}}{2\mu_T}$ and $\vec{v}^\perp \cdot \vec{q} = 0$.

We could understand some general features of the spectra independent of the velocity distributions. 
Given the exponential factor in the nuclear response function, all spectra fall off at large recoiling energies. In the heavy mediator limit, the spectrum of $\OO_1$ peaks at low recoil energy (or equivalently, small $v_{\min}$) since $g(v_{\min})$ is a monotonic decreasing function of $v_{\min}$. The differential rate of $\OO_{11}$ contains an additional factor, $v_{\min}^2$, which prefers larger $E_R$. Thus the spectrum peaks at a higher $E_R$, away from the detection threshold, due to a balance between $v_{\min}^2$ and $g(v_{\min})$. For $\OO_8$, the differential rate is a sum of two terms with opposite behavior: $v_{\min}^2 g(v_{\min})$ peaks at large recoiling energy while $h(v_{\min})$ peaks at threshold. Numerically it turns out that $h(v_{\min})$ dominates over $v_{\min}^2 g(v_{\min})$ so that the spectrum still peaks at low recoil energy. The situation is much simpler in the light mediator limit. The spectra for all operators peak at threshold due to the mediator propagator $1/q^2$.

We present the spectra for both light and heavy DM with masses at 12 and 50 GeV respectively in Fig.~\ref{fig:recoil_spectra_heavy} (in the heavy mediator limit) and Fig.~\ref{fig:recoil_spectra_light} (in the light mediator limit). In the plots, the cyan curves are based on SHM while the orange curves are based on the {\it Gaia} distribution with $\eta_{\rm sub}=0.42$. Based on the discussion above, we could understand further some details of the spectra for each velocity distribution.

In the heavy mediator case,
\begin{itemize}
\item For both $\OO_1$ and $\OO_{11}$, spectra based on {\it Gaia} peaks at a higher value at low recoiling energy and falls off faster compared to the one from SHM given the shapes of $g(v_{\min})$ shown in Fig.~\ref{subfig:g_vmin}. 
\item For $\OO_1$, the spectrum based on {\it Gaia} distribution is steeper for light DM than that for heavy DM. More specifically, the recoiling energy at which the differential rate from SHM becomes larger than that from {\it Gaia} for heavy DM (the cross-over of the two curves) is larger than that for light DM. This is because at $v_{\min} \sim 250$ km/s, the relative sizes of $g(v_{\min})$'s for {\it Gaia} distribution and SHM switches (Fig.~\ref{subfig:g_vmin}). For a given $v_{\min}$, the more heavy DM is, the larger $E_R$ it corresponds to, as one could see from Eq.~\eqref{eq:qv}. Similar argument could be used to explain the differences for spectral of light and heavy DM scattering through $\OO_{11}$. 
 \item For $\OO_8$, the spectra based on {\it Gaia} are always below those based on SHM since the scattering is mostly determined by $h(v_{\min})$. 
\end{itemize}

Similar features are present in the light mediator case (though they are less evident in Fig.~\ref{fig:recoil_spectra_light}): 
\begin{itemize}
\item For $\OO_{1}$ and $\OO_{11}$, the {\it Gaia} spectra are steeper than those of SHM. At low $E_R$, the differential rate based on the $\Gaia$ distribution is greater than that based on SHM while the opposite is true at higher $E_R$. Analogous to the heavy mediator case, the cross-over happens at a larger $E_R$ for heavy DM compared to light DM. 
\item For $\OO_{8}$, {\it Gaia}'s differential rate is always below the SHM's mainly due to the dominance of $h(v_{\min})$.
\end{itemize}

In general, the overall rate of light DM with mass $\lesssim 10$ GeV could be suppressed with a {\it Gaia} distribution compared to that of SHM. This leads to weaker constraints and poorer determinations of the light DM parameters using {\it Gaia} velocity distribution. On the other hand, for heavy DM, the relative sizes of scattering rate with either {\it Gaia} or SHM distributions depends on the type of the interaction. The scattering rate with {\it Gaia} could be enhanced when the associated velocity moment is $g(v_{\min})$ and the scattering rate is proportional to a non-negative power of $v_{\min}$ (or equivalently, $q$). This could lead to stronger constraints and better determinations of the DM parameters, e.g., when the model maps onto $\OO_1$ and $\OO_{11}$ with heavy mediators. For interactions associated with $h(v_{\min})$, however, the recoil rates assuming {\it Gaia} distribution are suppressed compared to that of SHM, resulting in a weaker constraint on the coupling. Lastly, the differences of the spectral shapes in the light mediator case, which depend on negative powers of $v_{\min}$ ($q$), with either {\it Gaia} or SHM distributions are small. These qualitative discussions based on the recoil spectra will indeed be confirmed with numerical computations in Sec.~\ref{sec:results}.

\section{Statistical framework}
\label{sec:method}
In this section, we introduce the statistical framework to study the effect of uncertainties in the DM velocity distribution while reconstructing particle physics parameters with DD data, followed by a short discussion on how to interpret its results with a concrete example.. As discussed previously in the introduction, for a given DD experiment, there is a three-fold degeneracy between the different classes of DM parameters. At the same time, from a statistical inference perspective, we can only access (a combination of) these parameters through experimental observables. For DM-nuclear interactions, these observables are simply the {\it overall rate} and the number of events per bin or the {\it shape}\ of the recoil spectrum (which may also include background events). The qualitative relationships between these observables and DM parameters of interest are summarized in table~\ref{tab:dd_summary}. 

\begin{table}[ht]
\centering
\begin{threeparttable}
	\begin{tabular}{| l | l  | l  | l  |}
	\hline
	 Type & \diagbox{Signal parameters}{Observables} & Total rate & Shape	\\ \hline \hline
\multirow{4}{*} {Particle and nuclear physics}& DM mass [$m_\chi$] & \ding{51} &  \ding{51} \\[3pt]
	& Couplings  [$c$'s] & \ding{51} & \ding{55}$^*$ \\[3pt]         
         & Mediator mass [$m_{\rm med}$] & \ding{51} & \ding{51}$^{**}$ \\[3pt] 
        & Form factor [${\cal F}$'s] & \ding{51} & \ding{51} \\[3pt]  \hline
\multirow{2}{*}{Astrophysics} & Local DM density [$\rho_\chi$] & \ding{51} & \ding{55}\\[3pt]
        & DM velocity distribution [$f(v)$] & \ding{51} & \ding{51} \\[2pt] 	
	\hline        
        \end{tabular}
		\begin{tablenotes}
	   \small
    \item $^*$ Exceptions occur when multiple relativisitc interactions are relevant, each giving rise to a different shape. Then varying couplings (non-uniformly) change the weights of each interaction. 
   \item $^{**}$ Applicable only for light mediators, {\it i.e} when $m_{\rm med}/q \lesssim 1$ as well as the case with multiple mediators. 
       \end{tablenotes}
\caption{Schematic summary of the relationship between DM signal parameters and experimental observables in a typical DD experiment.} \label{tab:dd_summary}
\end{threeparttable}
\end{table}

However, no statistically significant number of DM events have been detected at \textsc{Pico}~\cite{Amole:2015lsj}, \textsc{Lux}~\cite{Akerib:2016vxi}, \textsc{SuperCDMS}~\cite{Agnese:2017njq}, \textsc{PandaX-II}~\cite{Cui:2017nnn}, and \textsc{Xenon-1T}~\cite{Aprile:2018dbl}. These null results in turn can be used to obtain constaints on particle physics parameters, traditionally expressed through 90\% confidence level (CL) upper limit in the plane of dark matter mass and scattering cross section per nucleon. While studying upper limits is the most straightforward way to assess the impact of qualitatively different velocity distributions on extracting particle physics parameters with DD experiments~\cite{Necib:2018iwb, OHare:2018trr, Evans:2018bqy, Wu:2019nhd, Buckley:2019skk}, we could gain more information from forecasting the ability of next-generation experiments to reconstruct model parameters through pairwise comparison of neighboring points. We also note that several analyses in the literature have forecasted degeneracies between DM model parameter by analyzing mock data sets generated for various experimental configurations~\cite{McDermott:2011hx, Catena:2014epa, Gluscevic:2014vga, Gluscevic:2015sqa, Gelmini:2018ogy, Kahlhoefer:2017ddj}, but these methods rely on the choice of several pre-defined benchmark points and some of them could be computationally expensive.

A faster alternative for forecasting are techniques that rely on the so-called Asimov data set~\cite{Cowan:2010js}, an artificial data with no statistical fluctuations generated using the true parameter values of a model. Consider a $d$-dimensional parameter space, $\vec{\bm \theta} \in \Omega_{\mathcal{M}} \in \mathbb{R}^d$ for a given DM model $\mathcal{M}$, and the associated Asimov data set $D_A (\bm \theta)$ for each point. Given two model parameter points $\vec{\theta}_{1, 2} \in \Omega_{\mathcal{M}}$, we can construct a likelihood ratio test statistic (TS)~\cite{1933RSPTA.231..289N},
\beq \label{eq:ts0}
\text{TS} = -2 \, \loge \, \frac{\mathcal{L}_X(\mathcal{D}_A(\vec{\theta}_2)|\vec{\theta}_1)}{\mathcal{L}_X(\mathcal{D}_A(\vec{\theta}_2)|\vec{\theta}_2)} \approx \sum_{i, j=1}^{n_b} (\vec{\theta}_1 - \vec{\theta}_2)_{i} \, \widetilde{I}_{ij} \, (\vec{\theta}_1 - \vec{\theta}_2)_{j} \sim \chi^2_d \; ,
\eeq
which asymptotically has a $\chi^2$ distribution with $d$ degrees of freedom~\cite{Wilks:1938dza}. Here, $\tilde{I}$ is the profiled Fisher information matrix, and the summation runs from 1 to $n_b$, the number of data bins. While the TS is used to reject the null hypothesis that $\vec{\theta}_1$ and $\vec{\theta}_2$ are indistinguishable at the $(1 - \alpha)\%$ confidence level, Eq.~\ref{eq:ts0} can also be suitably modified to obtain sensitivity projections for future experiments in terms of 90/95\% CL upper limits.

Despite using the Asmiov data set, calculating the TS in Eq.~\eqref{eq:ts0} for model comparison of $N$ points in the parameter space can be expensive when $N$ is large. Thus, to facilitate fast, {\it benchmark-free} model comparison, refs.~\citep{Edwards:2017mnf, Edwards:2017kqw} introduced a novel method based on information theoretic techniques.\footnote{A \texttt{python} implementation of their results is available in the open source \texttt{swordfish} code: \url{https://github.com/cweniger/swordfish}, and a proof-of-concept application to future DD searches has been demonstrated in ref.~\citep{Edwards:2018lsl}.} Noting that the profiled Fisher information $\widetilde{I}$ transforms as a {\it metric} on the parameter space $\vec{\bm \theta}$, ref.~\cite{Edwards:2018lsl} mapped the parameter space into a higher-dimensional signal space and expressed the TS as a euclidean distance between two signals. More concretely, they used the embedding, $\vec{\bm \theta} \mapsto x(\vec{\bm \theta}) \in \mathbb{R}^{n_b}$, to transform the parameter space to the $n_b$-dimensional signal space with unit Fisher information matrix. After this transformation the TS can be written in terms of the appropriately named euclidean signal $x_i$,
\beq \label{eq:euclidean}
TS \approx || \vec{x}(\vec{\bm \theta}_1) - \vec{x}(\vec{\bm \theta}_2) ||^2. 
\eeq

Eq.~\eqref{eq:euclidean} is the main ingredient of our benchmark-free forcasting approach. In the language of this method, as long as the parameter space is sufficiently sampled, signal discrimination is only possible at the $(1 -\alpha)\%$ CL if the signals from two parameter points are at least a distance $r_\alpha (\mathcal{M})$ apart in the projected signal space. The distance, in turn, is related to the sampling distribution of the TS,
\beq \label{eq:radius}
r_\alpha (\mathcal{M})^2 \leq P_{\chi^2_d}^{-1} (1 - \alpha), 
\eeq 
where $P_{\chi^2_d}$ is the cumulative distribution function (CDF) of the $\chi^2$ distribution with $d$ degrees of freedom.

\begin{figure}[!ht]
\centering
 \begin{subfigure}[!h]{0.48\textwidth}
        \includegraphics[width=\textwidth]{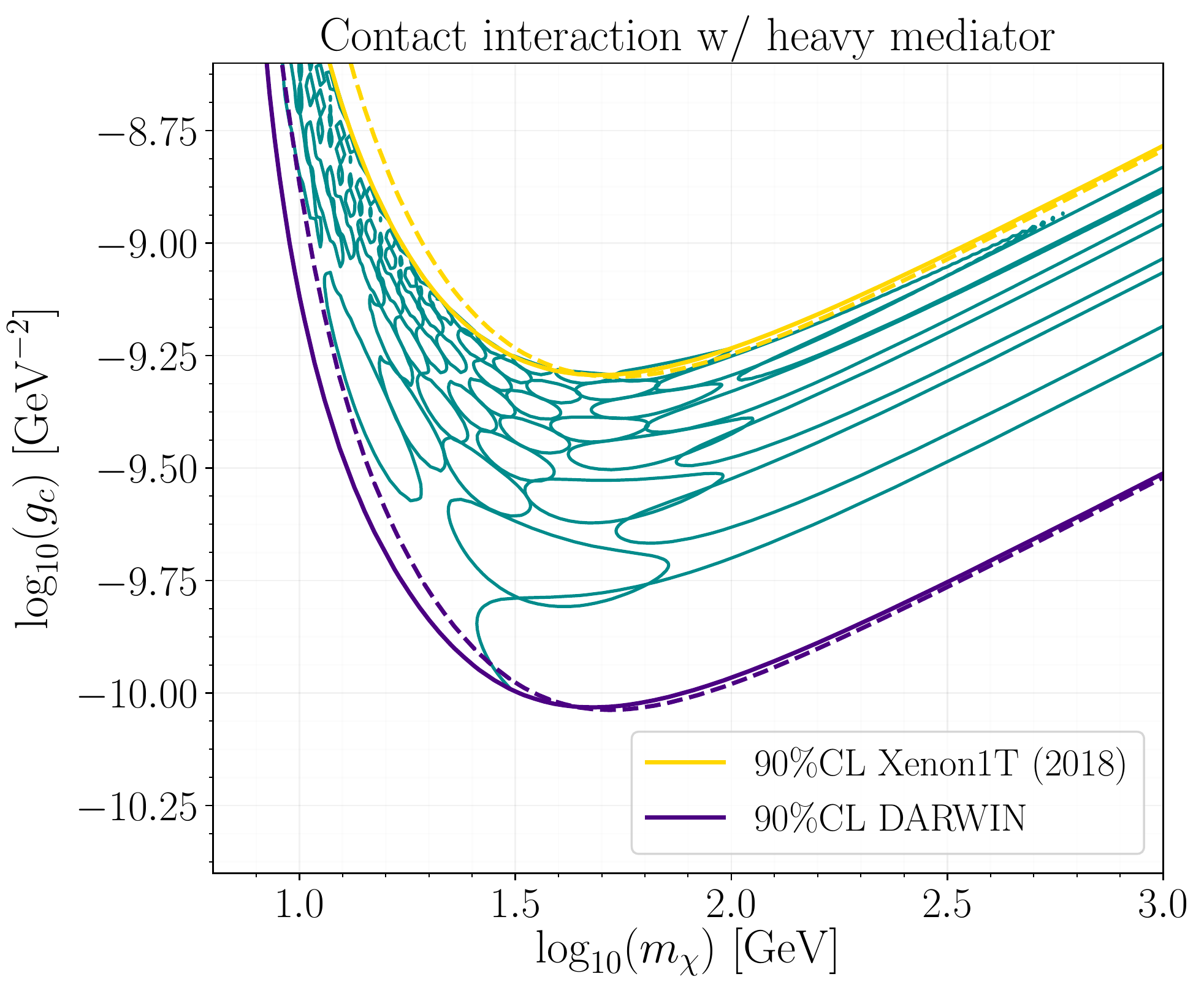}
    \end{subfigure}
    ~ \,
    \begin{subfigure}[!h]{0.48\textwidth}
        \includegraphics[width=\textwidth]{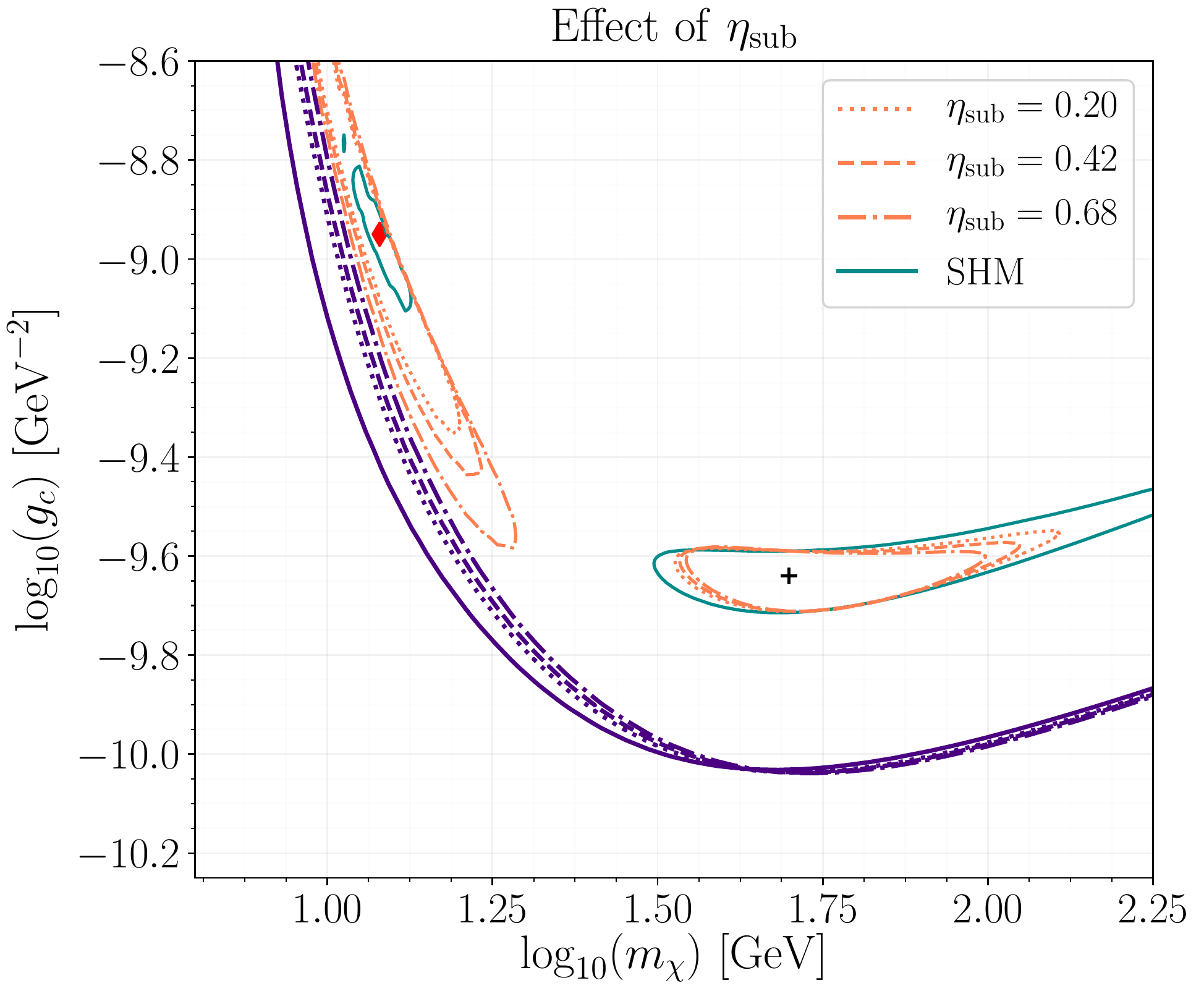}
    \end{subfigure}
\caption{Left: the `fishnet' plot obtained using the ES method is capable of illustrating the degeneracies between various pairs of signal parameters without recourse to computationally expensive MC simulations. The closed ellipses represent the usual 68\% CL contours in parameter space for arbitrary benchmark points at a \textsc{Darwin}-like experiment. Also indicated for reference are $90$\% CL upper limits following the latest \textsc{Xenon-1T} results (yellow) and projected upper limits for a \textsc{Darwin}-like experiment (indigo) assuming SHM (solid) and {\it Gaia} (dashed) velocity distributions. Right: 68\% CL contours and 90\% CL upper limits assuming a {\it Gaia} velocity distribution with different DM substructure fractions (dotted, dashed, and dot-dashed). These are shown alongside the SHM (solid) constraints to demonstrate that the dominant uncertainty is due to differences in the velocity distributions, and not the DM substructure fraction. }\label{fig:sum_plot}
\end{figure}

The above procedure can be understood very loosely as comparing two distributions, albeit incorporating the fact that they arise from the same likelihood function. We emphasize that there is no mock data generated at any stage of our analysis. The use of Asimov data set in eq.~\eqref{eq:euclidean} implies that the sampling distribution gives the median significance for two hypothetical data sets which have the parameter points $\vec{\theta}_{1, 2}$ as their maximum likelihood estimates. We illustrate the efficacy of this method in Fig.~\ref{fig:sum_plot}, the left panel of which shows the constraints in DM mass-coupling space for a model with contact interaction mediated by a heavy vector particle. The closed ellipses represent the usual 68\% CL contours in parameter space for arbitrary benchmark points at a next-generation \textsc{Darwin}-like liquid Xenon (LXe) experiment~\cite{Aalbers:2016jon} assuming SHM. These are obtained by constructing hyperspheres of radius, $r_\alpha$, in the Euclidean signal space, and back-projecting them to the parameter space using a lookup table for the embedding map. For a $\chi^2_d$ distribution with $d=2$, eq.~\eqref{eq:radius} implies that 68\% CL corresponds to a threshold value of $r_{0.32} =1.52$.

\section{Results}
\label{sec:results}
In this section, we combine the formulae and methodology presented in previous sections for studying how DM velocity distribution inferred from {\it Gaia} sausage could affect reconstruction of various DM particle physics parameters at next-generation DD experiments. For concreteness, we only consider a \textsc{Darwin}-like liquid Xenon (LXe) experiment and a complementary \textsc{DarkSide-20k}-like Argon experiment with both high~\cite{Aalseth:2017fik} and low mass~\cite{Agnes:2018ves} search programs (see Appendix~\ref{sec:nextgenexp} for more details). Unlike Sec.~\ref{sec:DMvelocity}, we present our results for the {\it benchmark} DM models listed in Table~\ref{table:SI_models} instead of individual operator in NREFT. While examining recoil spectrum of each operator is insightful, concrete models, especially those with well-motivated UV completions and/or distinct phenomenologies, enable an easy comparison of our results with those in the literature. In addition, since there could be non-trivial mapping between a model and NREFT, it may not be straightforward to find the sensitivity of DD to a full-fledged model by combining the sensitivity to individual NR operators. Yet we will still find the qualitative understanding developed in Sec.~\ref{sec:DMvelocity} a useful starting point for the results discussed here.

\subsection{DM Mass - coupling} \label{sec:mass_coup}
\begin{figure}[p]
    \centering
    \includegraphics[width=0.9\textwidth]{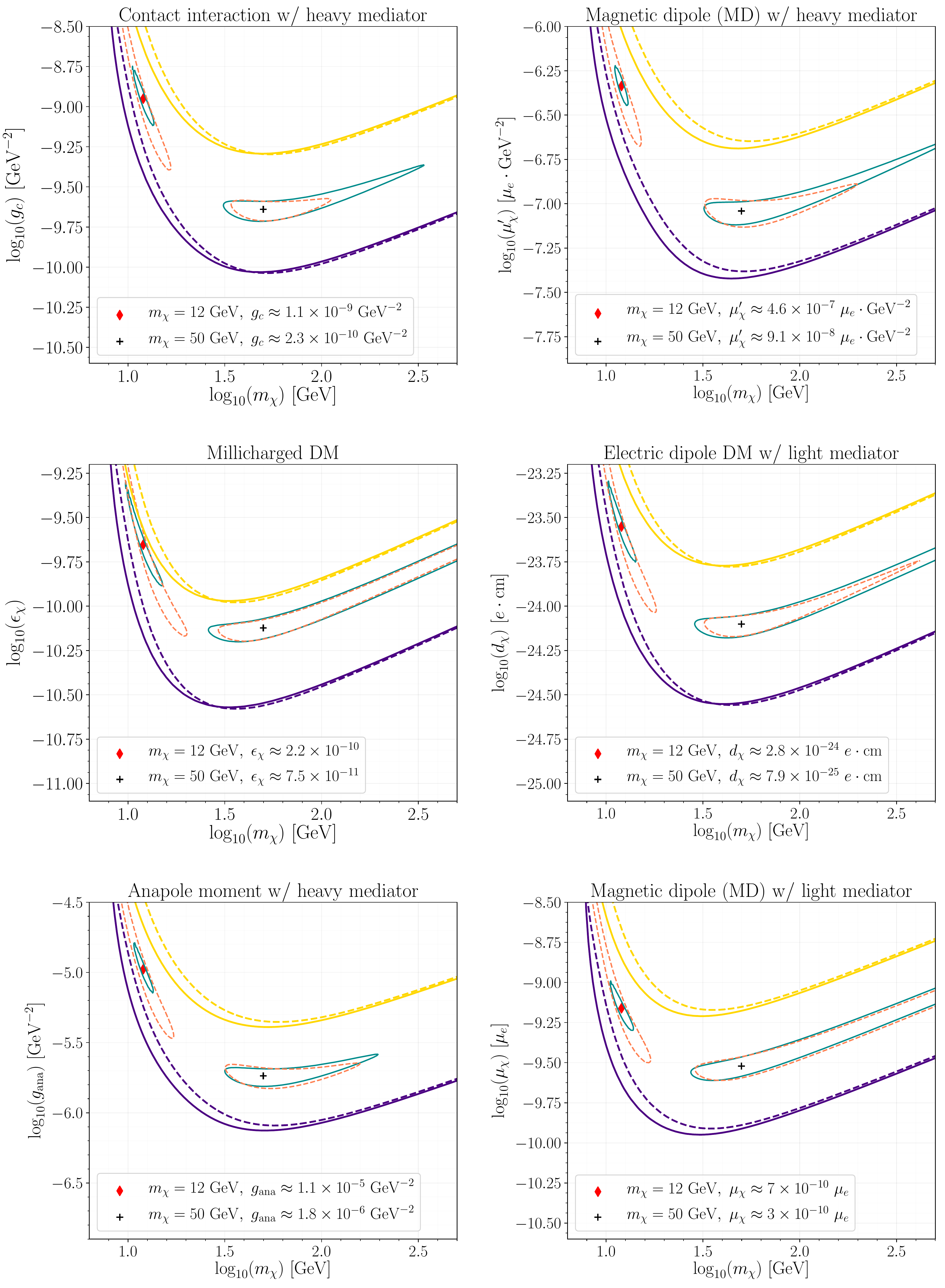}
    \caption{Constraints and forecasts in the DM coupling-mass plane for all the benchmark models in Table~\ref{table:SI_models} with varying $q^2$ and $v^2$ dependence. The $68$\% CL forecast contours for SHM (cyan, solid) and {\it Gaia} (orange, dashed) velocity distributions are shown for both light (red diamond, $m_\chi = 12$ GeV) and heavy (black cross, $m_\chi = 50$ GeV) DM. Also indicated for reference are $90$\% CL upper limits following the latest \textsc{Xenon-1T} results (yellow) and projected upper limits for a \textsc{Darwin}-like experiment (indigo) assuming SHM (solid) and {\it Gaia} (dashed) velocity distributions. The constraints for MD with heavy mediator are quoted in units of electron Bohr magneton, $\mu_e = \frac{e}{2m_e}$. }\label{fig:signals}
\end{figure}

We forecast the sensitivity of a next-generation LXe, \textsc{Darwin}-like, experiment to simultaneously resolve the DM mass and coupling for SHM and {\it Gaia} velocity distributions. Our first step is to investigate the effect of uncertainty in the DM substructure fraction $\eta_{\rm sub}$, estimated by ref.~\cite{Necib:2018igl} to be $\eta_{\rm sub} = 0.42^{+ 0.26}_{-0.22}$. We show, in the right panel of Fig.~\ref{fig:sum_plot}, the 68\% CL contours and 90\% CL upper limits corresponding to a {\it Gaia} velocity distribution with the median and $\pm 1\sigma$ DM substructure fractions. Comparing them with constraints for the SHM leads us to conclude: \emph{the primary effect in reconstructing DM model parameters arises due to the qualitative differences between the SHM and \Gaia velocity distributions, while the variation of the substructure fraction is only a subdominant effect beyond it.} Thus, for the rest of our analysis, we fix the DM substructure fraction to its median value, $\eta_{\rm sub} = 0.42$.\footnote{The ES method also provides a straightforward way to marginalize over the uncertainty in nuisance parameters, such as $\eta_{\rm sub}$, through the inclusion of a penalisation term; see Appendix A of ref.~\cite{Edwards:2018lsl} for more details.} We have also checked that varying $\eta_{\rm sub}$ in its $1\sigma$ range does not affect our discussion for DM interactions beyond the minimal contact interaction. 

In Fig.~\ref{fig:signals}, we show the $68$\% CL contours obtained using the ES method for four DM models with characteristic $E_R$ and velocity moment dependences (outlined in table~\ref{table:SI_models}), namely contact interaction, millicharged DM, DM with electric and magnetic dipole moments, anapole DM with a heavy mediator and magnetic dipole DM with a light mediator for SHM and {\it Gaia} velocity distributions. For reference, we also plot the $90$\% CL upper limits following the latest \textsc{Xenon-1T} results~\cite{Aprile:2018dbl} and projected upper limits for future LXe experiments using the {\it equivalent counts} method~\cite{Edwards:2017mnf, Edwards:2017kqw}. We constrain couplings instead of cross-sections (cf.~\cite{Gluscevic:2015sqa, Edwards:2018lsl}) as we don't integrate over the entire $E_R$ range to obtain the respective cross-sections for the DM models we consider here. Although, as illustrated in the left panel of Fig.~\ref{fig:sum_plot}, the ES method allows us to plot the degeneracy contour for any point in the parameter space, we show our results at two benchmark points corresponding to light ($m_\chi = 12$ GeV) and heavy ($m_\chi = 50$ GeV) DM for easier interpretability of our results. To ensure we are making an apples-to-apples comparison when studying the changes in constraints across models, we choose couplings such that the number of events is the same for each benchmark point with SHM. 

Before discussing the effect of DM velocity distribution, we explain the general behavior of constraints in DM coupling-mass space given in Fig.~\ref{fig:signals} in terms of the recoil spectra shape and the total event rate. For a given DM model, the recoil spectra for low mass DM peaks closer to threshold than for heavy DM. Moreover, the shape of the recoil spectra is degenerate only for a narrow range of masses, whereas a change in the total rate can be compensated by a wide range of couplings. As a result, in the light DM regime, we observe a large degeneracy in the coupling but a reasonable DM mass resolution. Conversely, for heavier DM, a DD experiment is more sensitive to the couplings but suffers from poor mass resolution, because the recoil spectra shape is degenerate for a wide range of DM mass. Thus, given a similar number of events, we can heuristically treat the light and heavy DM regimes as shape- and total rate-limited respectively. 

While these interpretations hold generally, certain qualitative details like the shapes and sizes of the contours can vary significantly between the SHM and {\it Gaia} velocity distributions. For instance, in the light DM regime, we observe that the contour size increases for all models with {\it Gaia} velocities, implying that LXe experiments have reduced sensitivity to {\it Gaia} distribution as compared to the SHM. This observation is consistent with the weakening of upper limits on SI cross-section for light DM first reported by ref.~\cite{Necib:2018iwb} with the latest \textsc{Xenon-1T} data. Our results also indicate an interesting effect of {\it Gaia} velocity distribution that has not been previously discussed in the literature: {\it depending on the model, there is a marginal improvement in the sensitivity of LXe experiments to heavy DM for {\it Gaia} when contrasted with the SHM}. 
More concretely, in the case of contact interaction, when dark matter mass is at 12 GeV, the resolution of coupling assuming {\it Gaia} distribution is reduced by a factor of 7 compared to SHM. On the other hand, when dark matter mass is at 50 GeV, the mass resolution is improved by a factor of 3 compared to SHM. 
We also note that for models with light mediators, the experimental sensitivity becomes poorer across the entire mass range irrespective of the DM velocity distribution.

These results can be understood, at least to leading order, in terms of the $E_R$ and velocity moment dependences of each model. We start with the observation that contours of the {\it Gaia} velocity distribution are less constraining than those of SHM at low DM masses for all models we consider here. These models have a leading order DM velocity moment that scales as $g(v_{\min})$ and/or $h(v_{\min})$ suppressed or enhanced by additional powers of $E_R$. Since light DM corresponds to a high $v_{\rm min}$ for a heavy target like xenon, only the tail of the $g(v_{\min})$ and/or $h(v_{\min})$ distribution (Fig.~\ref{fig:f_v_int}) contributes to the recoil rate, where the SHM curves always dominate over the {\it Gaia} ones. 

For heavy DM, on the other hand, varying DM mass could lead to sharper changes in the recoil spectra shapes with the {\it Gaia} distribution as compared to the SHM. Thus, there is an improvement in the sensitivity to DM mass as evidenced by the shrinking of 68\% CL forecast contours in the mass direction in Fig.~\ref{fig:signals}. Moreover, as shown in the top row, this effect is most apparent for models with a heavy mediator, or equivalently for non-negative powers of $E_R$. We also note that there could be some subtle difference, e.g., between models with contact interaction and magnetic dipole interaction. For contact interaction, the upper limits for DM with mass above 50 GeV are slightly tightened with the {\it Gaia} distribution. This is due to an enhanced recoil rate contributed by {\it Gaia}'s larger $g(v_{\min})$ at $v_{\min} {\sim} 150$-$200$ km/s as compared to SHM. Whereas, in the case of magnetic dipole DM, the {\it Gaia} upper limits for DM with mass above 50 GeV are slightly weakened, since the scattering rate of magnetic dipole DM scales as $E_R^2 g(v_{\min})+ E_R h(v_{\min})$. The second term proportional to $h(v_{\min})$ leads to a small reduction in the overall recoil rate with the {\it Gaia} distribution, as compared to the SHM one. Meanwhile, the positive powers of $E_R$ for magnetic dipole interaction result in an enhanced sensitivity of forecasts using the {\it Gaia} distribution. This is an interesting example where despite the reduction in the total number of events, the sensitivity actually improves with the {\it Gaia} distribution! 

In case of light mediators, the inverse powers of $E_R$ make the recoil spectra peak sharply as $E_R \to 0$. Yet for a finite threshold $E_R \approx 5$ keV, the \textsc{Darwin}-like experiment is only sensitive to the tail resulting in highly degenerate recoil spectra for different velocity distributions. This leads to poorer experimental sensitivity for both SHM and {\it Gaia} velocity distributions across all DM masses for models with light mediators, compared to the contact interaction model with a heavy mediator. In this case, lowering the detection threshold could improve the sensitivity to light mediators, making another physical case for the low threshold frontier.


\subsection{Mediator - DM mass} \label{sec:med_dm}

\begin{figure}[!ht]
\centering
\includegraphics[width=1.01\textwidth]{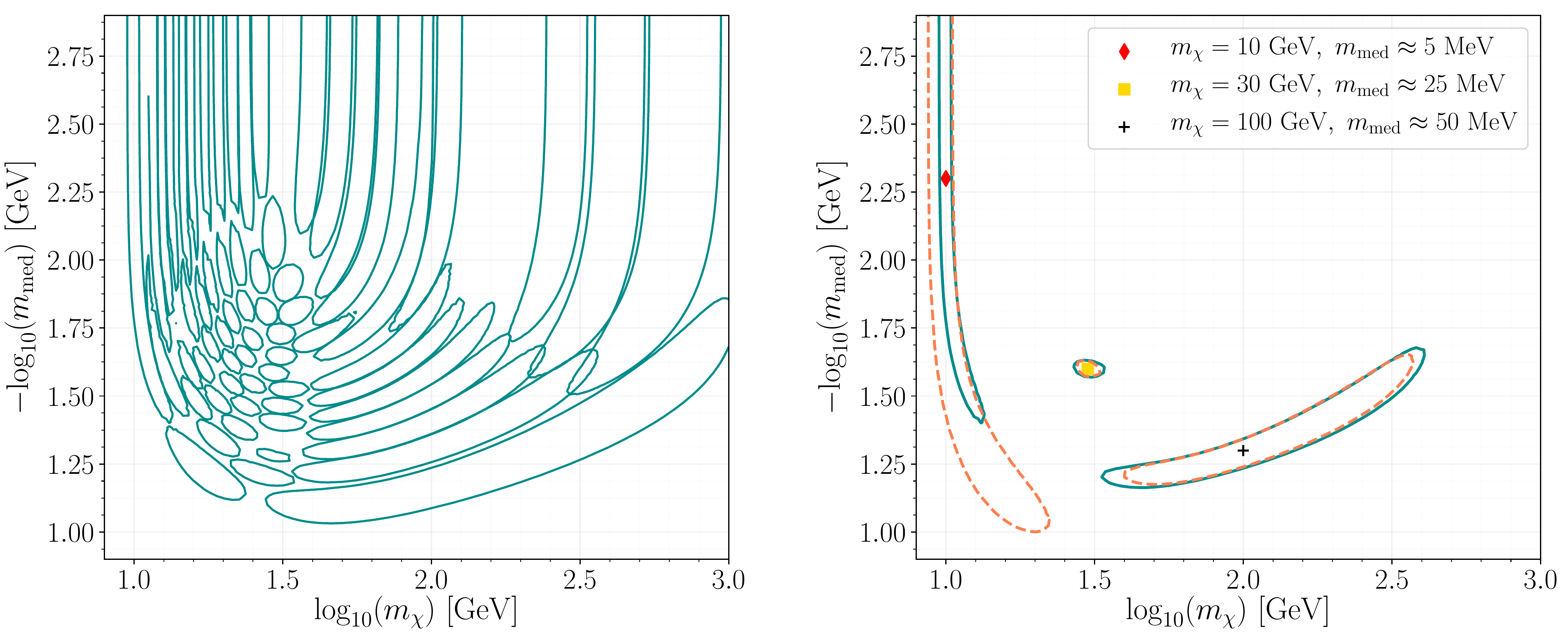}
 \caption{Forecasts in the $m_{\rm med}^{-1}-m_\chi$ plane for a DM with contact interaction mediated by a light scalar particle. Left: lattice of non-overlapping $68$\% CL forecast contours for the SHM. Right: $68$\% CL contours for SHM (cyan, solid) and {\it Gaia} (orange, dashed) velocity distributions shown for three benchmark points corresponding to light (red diamond), intermediate (yellow square), and heavy (black cross) DM with mediator masses in the $1$-$50$ MeV range.}\label{fig:medvmass}
\end{figure}

DD experiments are best suited to constrain the mediator mass $m_{\rm med}$ when it is at the same order of the momentum exchange, i.e. ${m_{\rm med} \sim q \sim \mathcal{O} (10\, {\rm MeV})}$~\cite{Kahlhoefer:2017ddj, DelNobile:2015uua, Aprile:2019dbj}. We use the ES method to simultaneously constrain the mediator and DM masses at a \textsc{Darwin}-like experiment for a fixed coupling. We also study the effect of the DM velocity distribution on the forecast. For simplicity, we only consider DM with contact interaction mediated by a light scalar particle and fix the DM coupling to be the same for all benchmark points. 

In the left panel of Fig.~\ref{fig:medvmass}, we show the $68$\% CL forecast contours for SHM in the $m_{\rm med}^{-1}-m_\chi$ plane. Broadly, the structure of these contours resembles the fishnet plot in the DM mass-coupling plane shown in the left panel of Fig.~\ref{fig:sum_plot}. On closer inspection, however, we can roughly delineate three regimes of sensitivity in this parameter space. The upper part of the plots is the light mediator ($m_{\rm med} \ll q$) regime where the propagator squared simply scales as ${\sim}1/q^{4}$. In this case, the $m_{\rm med}$ dependence drops out, and DD experiment is insensitive to the mediator mass. Next, we consider the heavy DM regime in the lower right part of the plots, where we have chosen the mediator mass such that $m_{\rm med} \lesssim q$. We find that the degeneracy in the recoil spectra in this limit is due to the DM mass, and any change in the mediator mass effectively acts as a rescaling of the overall coupling. Lastly, for intermediate DM mass and $m_{\rm med} \lesssim q$, our \textsc{Darwin}-like experiment can precisely reconstruct both mediator and DM masses primarily due to the high number of signal events (a factor of a few greater than the other benchmark points) in this regime.

We also illustrate the differences between SHM and {\it Gaia} velocity distributions in the right panel of Fig.~\ref{fig:medvmass} by plotting the $68$\% CL contours for three benchmark points: light ($m_\chi = 10$ GeV), intermediate ($m_\chi = 30$ GeV), and heavy ($m_\chi = 100$ GeV) DM with mediator masses in the $1$-$50$ MeV range. The constraints for SHM and \Gaia velocity distributions at the light and heavy DM benchmark points broadly follow the trend discussed for Fig.~\ref{fig:signals} in the previous section, while the difference between them is negligible in the intermediate DM mass regime.

\subsection{Model discrimination} \label{sec:model_dis}
\begin{figure}[!htb]
    \centering
        \includegraphics[width=1.\textwidth]{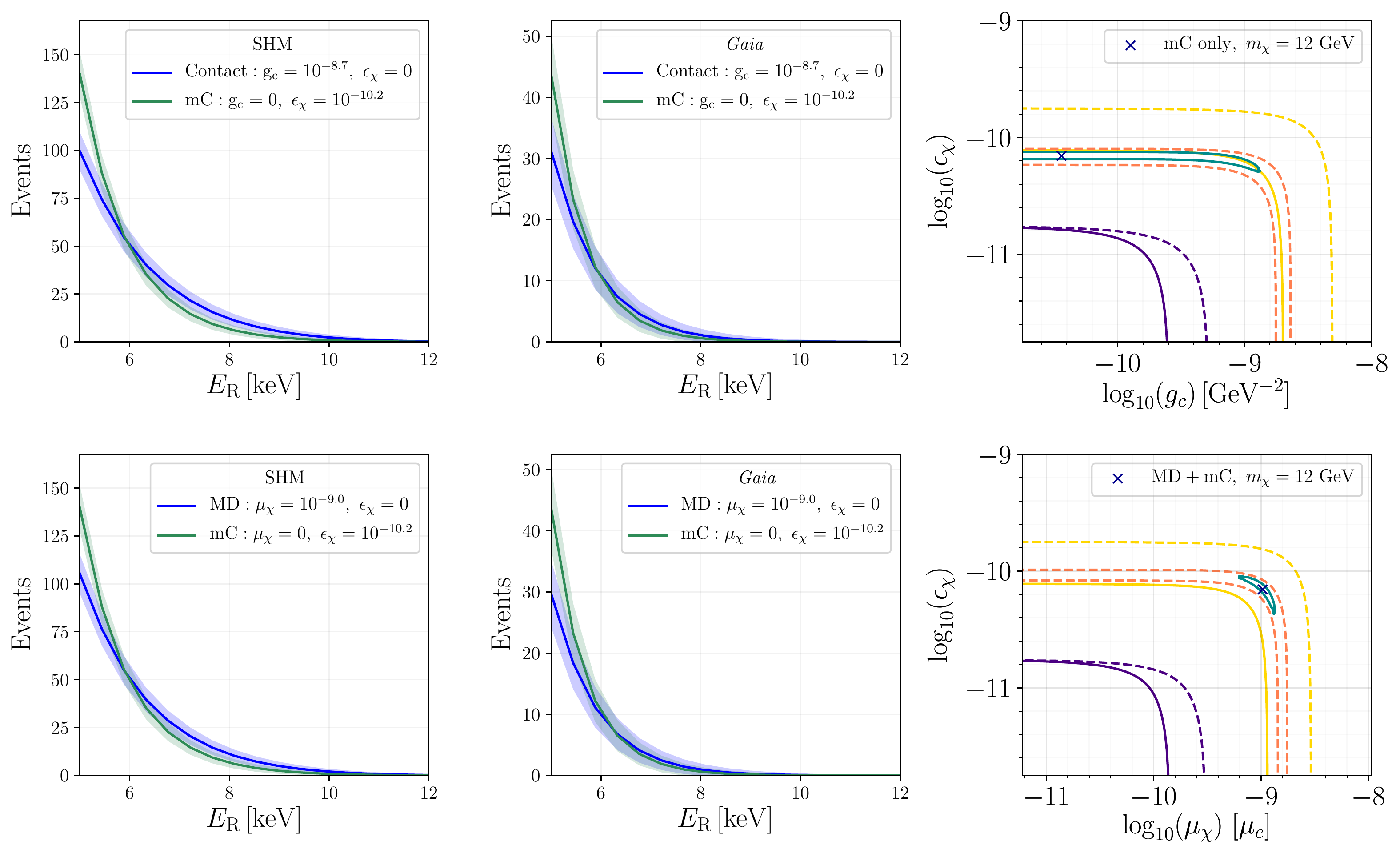}
    \caption{Forecasts for model discrimination in case of a 12 GeV DM at a \textsc{Darwin}-like experiment. The first two columns are recoil spectra for the two models under comparison with either SHM or {\it Gaia} velocity distributions, where the light-colored bands indicate the $1\sigma$ Poisson uncertainties. The third column shows the 68\% CL forecast contours for SHM (cyan) and {\it Gaia} (orange) in the coupling-coupling space for two comparison models: (first row) millicharge with a light mediator and contact interaction with a heavy mediator, (second row) millicharge and magnetic dipole both with a light mediator (SM photon). We also include $90$\% CL upper limits from the latest \textsc{Xenon-1T} results (yellow) and projected upper limits for a \textsc{Darwin}-like experiment (indigo) assuming SHM (solid) and {\it Gaia} (dashed) velocity distributions.}\label{fig:modelvmodel_L}
\end{figure}

\begin{figure}[!htb]
    \centering
        \includegraphics[width=1.\textwidth]{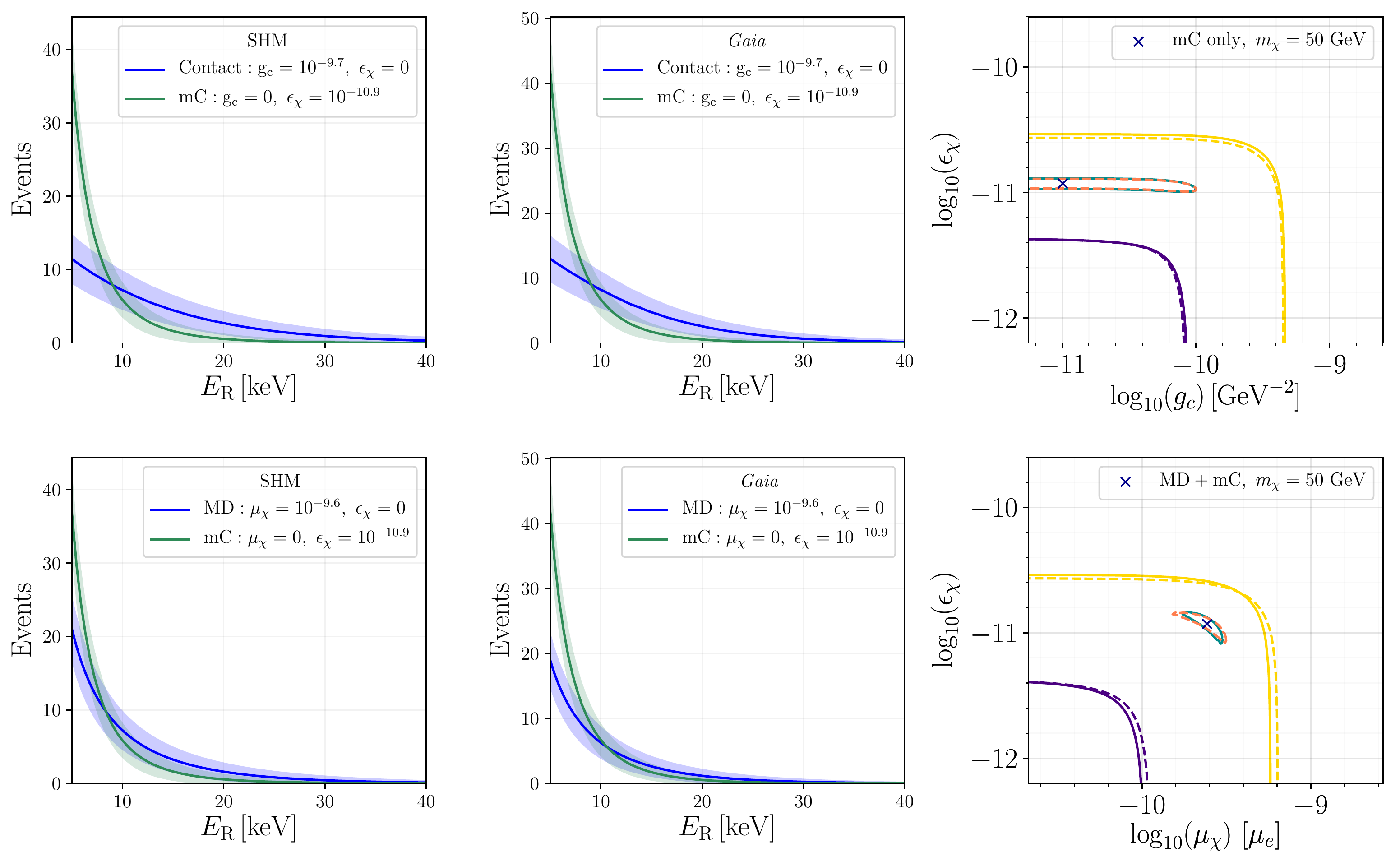}
    \caption{A similar plot as Fig.~\ref{fig:modelvmodel_L}, showing the same set of model pairs comparison, except that the DM mass is 50 GeV. }\label{fig:modelvmodel_H}
\end{figure}

In presence of a positive signal at a future DD experiment, one of the most important goals is to determine the type of DM-nuclear interaction and discriminate between different model candidates. To demonstrate the model selection, we postulate a scenario in which there are two candidate models of interest. We parameterize our model as the sum of a pair of interactions,
\beq\label{eq:modelvsmodel}
\mathcal{L}_{\rm int} = \frac{c_a\OO_a}{q^2+m^2_{{\rm med};a}} + \frac{c_b\OO_b}{q^2+m^2_{{\rm med};b}}.
\eeq
While holding $m_\chi$ and $m_{\rm med}$ fixed, we sample different values of $(c_a, c_b)$ and test how well a given recoil spectrum shape can determine the model parameters at 68\% confidence level. We test the two pairs of models: 1) $a,b=$ contact interaction, millicharge, and 2) $a,b=$ magnetic DM with light mediator and millicharge respectively. We present the results in Fig.~\ref{fig:modelvmodel_L} for light DM with mass at 12 GeV and Fig.~\ref{fig:modelvmodel_H} for heavy DM with mass at 50 GeV. These results are all based on a \textsc{Darwin}-like LXe experiment. 

In Fig.~\ref{fig:modelvmodel_L}, we present 68\% CL forecast contours in the coupling-coupling space for a 12 GeV DM particle. In the top row, we have a millicharged DM giving rise to an experimental signal and we want to test whether we would confuse it with the simplest contact interaction as both interactions lead to spectra peaking at experimental threshold. This scenario is equivalent to setting $c_a \neq 0,  c_b = 0$ in eq.~\eqref{eq:modelvsmodel}, where $c_a = \epsilon_\chi e$ for the millicharge model and $c_b = g_c$ for the contact interaction. In the bottom row, we have a DM particle with both magnetic dipole moment and millicharge interacting with the nucleus through the SM photon contributing with comparable rates, where $c_a = e\mu_\chi/2$ and $c_b=\epsilon_\chi e$. We want to test how well we could constrain the two relevant electromagnetic moments.

From the third column, one could see that while it is possible to reconstruct the model parameters (with large uncertainties) assuming SHM, the discrimination power is entirely lost with the {\it Gaia} distribution. The millicharged light DM could be misidentified as a light DM with simple contact interaction at 68\% C.L, as shown in the last plot in the top row. DM with both millicharge and magnetic dipole moment could not be distinguished from DM with only one of them, as shown in the last plot in the bottom row. This result could be understood from the recoil spectra shown in the first two columns, in which we fix the couplings of different DM models to give the same event numbers with SHM. Comparing the spectra based on the {\it Gaia} velocity distribution with those from SHM, one find that with {\it Gaia} distribution: {\it i)} the total number of events is significantly lower, which, in turn, increases the Poisson uncertainty; {\it ii)} the spectral shapes of different models, especially the tails of the distributions, are more degenerate. This is consistent with what we find in Sec.~\ref{sec:DMvelocity} using NREFT. 

Results of model discrimination for heavy DM with mass at 50 GeV are presented in Fig.~\ref{fig:modelvmodel_H}. In this case, we find that reconstructing the couplings is more accurate as compared to the light DM case, independent of the model combinations. We could almost distinguish different models or different combinations of model parameters equally well for both SHM and the {\it Gaia} distribution. The main reason is that the total number of events and the spectral shape do not change much when the velocity distribution varies. 

\subsection{Combining forecasts for different targets} \label{sec:targets}

The idea of combining different targets for a more accurate identification of momentum dependence of DM interactions, or for precise reconstruction of DM mass and model parameters is well-documented in the literature~\cite{Pato:2010zk, McDermott:2011hx, Cherry:2014wia, Gluscevic:2015sqa, Bozorgnia:2018jep}. We revisit this idea by using the ES method to forecast the sensitivity of two complementary next-generation experiments with different targets. In particular, we focus on a \textsc{Darwin}-like and a \textsc{DarkSide-20k}-like experiment with xenon and argon targets respectively. Besides forecasting for high mass DM as officially proposed by the \textsc{DarkSide-20k} collaboration~\cite{Aalseth:2017fik}, we also include results for a dedicated low DM mass search with a low threshold configuration similar to ref.~\cite{Agnes:2018ves}.

\begin{figure}[!htb]
    \centering
 \includegraphics[width=\textwidth]{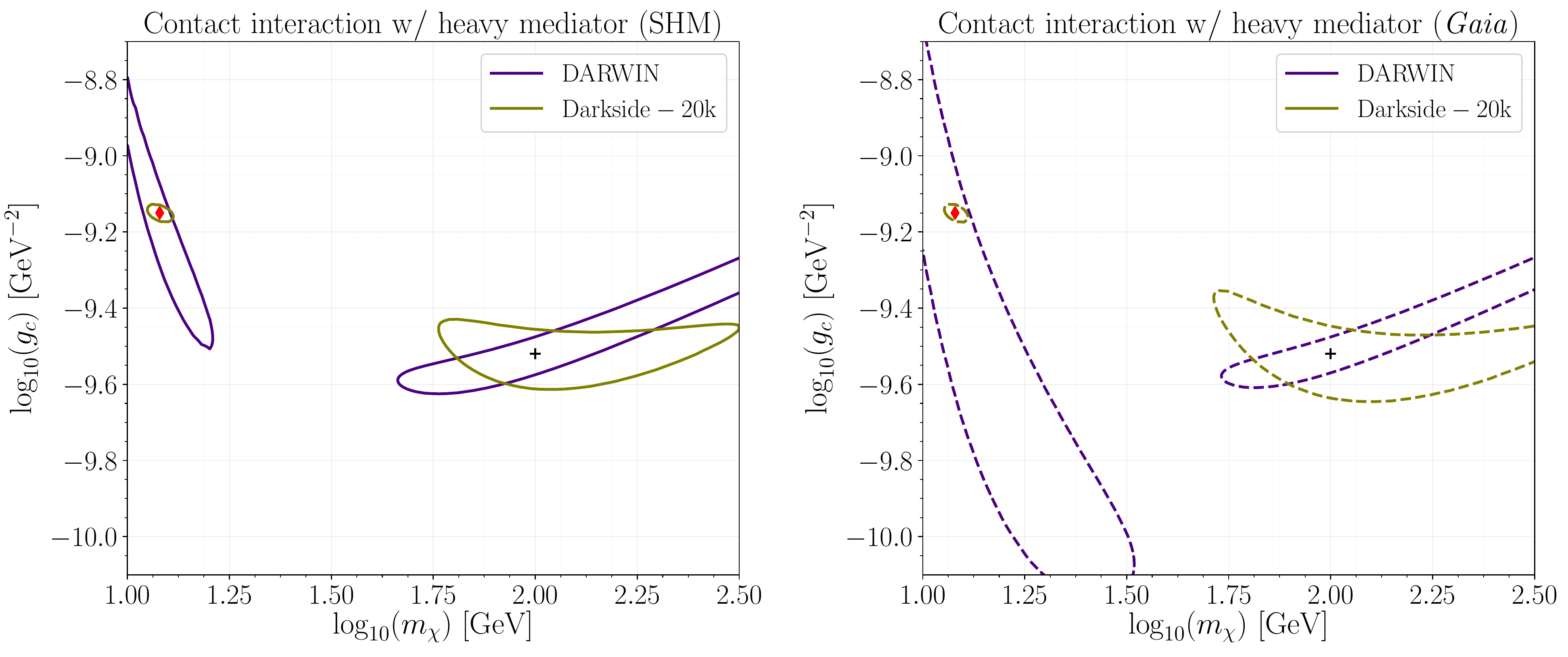}    
 \caption{Forecasts in the DM coupling-mass plane for DM contact interaction assuming SHM (left) and {\it Gaia} (right) velocity distributions. Unlike Fig.~\ref{fig:signals}, the $68$\% CL contours here represent forecasts for two complementary next-generation experiments: a \textsc{Darwin}-like experiment (indigo) and a \textsc{DarkSide-20k}-like experiment (olive) with xenon and argon targets respectively. Also indicated for reference are two benchmark points for light (red diamond, $m_\chi = 12$ GeV) and heavy (black cross, $m_\chi = 100$ GeV) DM.}\label{fig:darwin_ds20k_h}
\end{figure}

For simplicity, we consider the contact interaction as an example. 
Fig.~\ref{fig:darwin_ds20k_h} shows the $68$\% CL contours in the DM mass-coupling space for each of these experiments at two benchmark points corresponding to light ($m_\chi = 12$ GeV) and heavy ($m_\chi = 50$ GeV) DM. We find that combining forecasts from both xenon and argon targets could dramatically improve the sensitivity for low mass DM and, to a lesser extent, high mass DM. This result is independent of the model for DM velocity distribution. 
The extremely high resolution for low mass DM is due to the low threshold version of a \textsc{DarkSide-20k}-like experiment as listed in table~\ref{tab:futexp} of Appendix~\ref{sec:nextgenexp}. An advantage of using a lighter target like argon is that low DM masses correspond to much lower values of $v_{\rm min}$ as compared to xenon. This implies that the sensitivity forecast for a \textsc{DarkSide-20k}-like experiment are largely unaffected by the suppressed {\it Gaia} $g(v_{\rm min})$ distribution at large $v_{\rm min}$.

\section{Conclusions and Outlook}
\label{sec:con}

The new insight into the substructure of MW's DM distribution provided by the {\it Gaia} survey forces us to move away from the simplest SHM and to re-evaluate astrophysical uncertainties in DD experiments. In this paper, we investigate the effect of {\it Gaia} Sausage, one of the most established and representative substructures, on interpreting DD data for different DM models. We demonstrate that the new {\it Gaia} velocity distribution could result in potentially large modifications of both the overall scattering rate and the recoil spectral shape. Given the limited information from existing data sets, we focus on how the {\it Gaia} velocity distribution could affect forecasting at the next generation DD experiments with the euclideanized signal method.\footnote{Current experiments perform the full event-by-event likelihood analysis, e.g., as described in refs.~\cite{Aprile:2011hx, Aprile:2019dme}, which cannot be easily reproduced by researchers outside the collaboration. We hope that highlighting the importance of spectral shape information could motivate the experimental collaborations to release the full three dimensional likelihood function for reconstructed DM events in case of a positive detection.}

We study the sensitivity of DD experiments to different combinations of model parameters for representative DM benchmark models and its potential to distinguish different DM models given the {\it Gaia} distribution. We summarize our main findings below:
\begin{itemize}
\item While there is still uncertainty in the fraction of DM in the substructure, the primary effect in reconstructing DM model parameters is due to the qualitative differences between the shapes of {\it Gaia} and SHM distributions, which are independent of the precisie value of $\eta_{\rm sub}$. 
\item For light DM with mass at or below 10 GeV, the {\it Gaia} velocity distribution leads to a significantly weakened constraint for all the models we consider. Moreover, it poses a serious challenge for identifying the DM interaction and determining its strength assuming discovery at a \textsc{Darwin}-like experiment with threshold $E_R \approx 5$ keV. 
\item On the contrary, for heavy DM with mass above ${\sim 30}$ GeV, there could be a (moderate) improvement in the sensitivity of the next generation DD experiments when the mediator is heavy. 
\item Moreover, our results show that for positive (negative) powers of $E_R$ in the DM model, the sensitivity of a \textsc{Darwin}-like experiment is improved (worsened) irrespective of the DM mass and velocity distribution.
\item The additional challenge in probing light DM due to the {\it Gaia} distribution could be overcome using complementary experiments with lighter targets and lower thresholds.  
\end{itemize}

As mentioned in the Introduction, our work, along with several others, consists the early stages of a larger program to determine the DM phase space distribution, and assess its impact on various terrestrial experiments searching for DM in the {\it Gaia} era. In this paper, we restrict ourselves to models with leading order SI elastic scattering for DM with mass above GeV. It will also be of interest to extend the work to models with leading order spin-dependent and/or inelastic scattering. We intend to explore how the substructures discovered using {\it Gaia} data affect the new DD experiments probing DM-electron scattering in a future publication.

\begin{acknowledgments}
We thank Lina Necib for useful correspondence. We are also grateful to Thomas Edwards and Rick Gaitskell for providing useful feedback on the manuscript. The results in this work were computed using the following open-source software: \texttt{swordfish}~\cite{Edwards:2017kqw}, \texttt{IPython}~\cite{Perez:2007emg}, \texttt{matplotlib}~\cite{Hunter:2007ouj}, \texttt{scipy}~\cite{jones2001scipy}, and \texttt{numpy}~\cite{vanderWalt:2011bqk}. JF thanks KITP at UCSB (supported by the National Science Foundation under Grant No. NSF PHY-1748958) at which part of the work was carried out. JB, JF and JL are supported by the DOE grant DE-SC-0010010 and NASA grant 80NSSC18K1010.
\end{acknowledgments}

\appendix

\section{DD basics} \label{sec:dd_basics}
The rate of DM scattering off nucleus in a DD experiment is given by $R = n_\chi \left<\sigma v_\chi \right>$, where $n_\chi$ and $v_\chi$ are the local DM number density and speed relative to the Earth respectively, $\sigma$ is the scattering cross section and $\left< \ldots \right>$ indicates an average over the local DM velocity distribution. The differential recoil rate for a target $T$ per unit recoiling energy can be written as,
\beqa\label{eq:rate_short}
\frac{d R}{d E_{R}} &=& \xi_T N_T \frac{\rho_\chi}{m_\chi} \,  \int_{v_{\rm min}}^{v_{\rm esc}} d^3 v \, v \tilde{f}(\vec{v}) \frac{d \sigma}{d E_R}(v, E_R) \no \\
&=& \frac{\xi_T}{32\pi m_\chi^2m_T^2}\, \times \underbrace{\frac{\rho_\chi}{m_\chi} \, \int_{v_{\rm min}}^{v_{\rm esc}} d^3 v \, \frac{\tilde{f}(\vec{v})}{v}}_{{\rm astrophysics}} \times \underbrace{ \frac{1}{(2J+1)(2J_\chi+1)} \sum_{\rm spins} \left|\mathcal{M}\right|^2}_{{\rm particle/nuclear \ physics}},
\eeqa
where in the second line, we use 
\beq
\frac{d\sigma}{dE_R} = \frac{1}{32 \pi v^2} \frac{\overline{\left|\mathcal{M}\right|^2}}{m_\chi^2 m_T} , \quad {\rm with} \; \overline{\left|\mathcal{M}\right|^2} = \frac{1}{(2J+1)(2J_\chi+1)} \sum_{\rm spins} \left|\mathcal{M}\right|^2,
\eeq
and group together different factors by the main type of physics they rely on. In the master equation for the differential rate, $\xi_T$ is the mass fraction for each type of target nucleus $T$ (the detector could be composed of different nuclides), $m_T$ is the target nucleus mass and $N_T = 1/m_T$ is number of scattering centers per unit mass. $\rho_\chi$ is the local DM density in the solar system, which we take to be $\rho_\chi= 0.4~{\rm GeV/cm}^3$~\cite{Silverwood2017, Buch:2018qdr, 2019JCAP...10..037D}. While there could be an $\mathcal{O}(1)$ uncertainty in the determination of $\rho_\chi$ due to non-equilibrium effects in the dynamical modeling of the MW, we can always absorb it into the overall normalization of the recoil rate. In other words, local DM density only affects the overall rate but not the recoil shape. Thus in our paper, we will ignore uncertainty in the local DM density, focusing instead on the more interesting effects from varying the velocity distributions. $\tilde{f}(\vec{v})$ is the local DM velocity distribution in the Earth frame. The velocity integration range is bounded from above by the escape velocity of DM particles $v_{\rm esc}$. The minimal velocity for DM to scatter with recoiling energy $E_R$, in the case of elastic scattering, is 
\beq\label{eq:vmin}
v_{\rm min} = \sqrt{\frac{m_T E_R}{2 \mu_T^2}},
\eeq
where $\mu_T$ is the DM-nucleus reduced mass. Lastly, in the part that depends on particle and nuclear physics, $J$ and $J_\chi$ are nuclear and DM spins respectively. $\overline{\left|\mathcal{M}\right|^2}$ is the scattering matrix element squared averaged over $2J_\chi + 1$ and $2J+ 1$ initial DM and nuclear spins, and summed over the final spins. Note that the recoiling spectrum depends on the detector material in multiple ways. For example, in the velocity integration, $v_{\rm min}$ depends on the target nucleus mass and the matrix element depends on the nuclear form factor, which is determined by the type of the target. 

In terms of the form factors, the spin-averaged amplitude squared is then
\beq
\overline{\left|{\cal M} \right|^2} = 16 m_T^2 m_\chi^2  \sum_{\substack{i,j=1}}^{12}\sum_{\substack{N,N'=n,p}}\frac{c^{(N)}_i  c^{(N')}_j}{(q^2 + m_{{\rm med};i}^2)(q^2 + m_{{\rm med};j}^2)}\FF^{(N,N')}_{i,j}(q^2, v^2). 
\eeq

\section{Nuclear form factors in NREFT}
\label{sec:NREFT_operator}
An important ingredient for calculating the recoil rate is the form factor, $\mathcal{F}_{a,b}^{(N,N')} \sim \OO_a^{(N)}\times\OO_b^{(N')}$, that encodes nuclear response functions. Ref.~\cite{Fitzpatrick:2012ix} showed that the complete basis of NREFT operators corresponds to six different types of nuclear response functions $\widetilde{F}_i^{(N,N')}$, where $i \in \{M, \Delta, \Sigma', \Sigma'', \tilde{\Phi'}, \Phi''\}$ and the superscripts $N, N' = n,p$ indicates the type of nucleon. Listed below are the nuclear form factors for the relevant operators in terms of the response functions,
\begin{align}
& \mathcal{F}_{1,1}^{(N,N')} = \widetilde{F}_M^{(N,N')},  \\ 
& \mathcal{F}_{4,4}^{(N,N')} = \frac{1}{16}(\widetilde{F}_{\Sigma'}^{(N,N')}  + \widetilde{F}_{\Sigma''}^{(N,N')} ), \\
& \mathcal{F}_{5,5}^{(N,N')} = \frac{q^2}{4}\Big(\big(v^2-\frac{q^2}{4\mu_T^2}\big)\widetilde{F}_M^{(N,N')} +\frac{q^2}{m_N^2}\widetilde{F}_\Delta^{(N,N')} \Big),\\
& \mathcal{F}_{6,6}^{(N,N')} = \frac{q^4}{16}\widetilde{F}_{\Sigma''}^{(N,N')},  \\
& \mathcal{F}_{8,8}^{(N,N')} =\frac{1}{4}\Big(\big(v^2-\frac{q^2}{4\mu_T^2}\big)\widetilde{F}_M^{(N,N')}  +\frac{q^2}{m_N^2}\widetilde{F}_\Delta^{(N,N')} \Big),  \\
& \mathcal{F}_{9,9}^{(N,N')} = \frac{q^2}{16}\widetilde{F}_{\Sigma'}^{(N,N')}, \\ 
& \mathcal{F}_{11,11}^{(N,N')} = \frac{q^2}{4}\widetilde{F}_M^{(N,N')}, \\
& \mathcal{F}_{4,5}^{(N,N')} = \frac{q^2}{8m_N}\widetilde{F}_{\Sigma',\Delta}^{(N,N')},   \\
& \mathcal{F}_{4,6}^{(N,N')} = \frac{q^2}{16}\widetilde{F}_{\Sigma''}^{(N,N')}, \\
& \mathcal{F}_{8,9}^{(N,N')} = \frac{q^2}{8m_N} \widetilde{F}_{\Sigma',\Delta}^{(N,N')},
\end{align}
where the subscripts indicate the NREFT operator(s). The response functions $\widetilde{F}_i^{(N,N')}$ for the nuclides relevant in our analysis (see Appendix~\ref{sec:nextgenexp}) have been adopted from Appendix A.3 of ref.~\cite{Fitzpatrick:2012ix}.

\section{Some benchmark models}
\label{sec:models}
In this appendix, we present a few more details of benchmark models listed in Table~\ref{table:SI_models}. 

\noindent \textbf{Heavy gauge boson mediator:} In this case, quarks and DM are both charged under a broken gauge symmetry with a heavy gauge boson. The interactions are given by,
 \beq
\mathcal{L} \supset g_\chi \bar{\chi}\gamma_\mu\chi Z^{\prime \mu} + g_q \bar{q}\gamma_\mu q Z^{\prime \mu},
\eeq 
where $Z^\prime$ is the heavy gauge boson that could be integrated out, giving rise to a four fermion contact operator. 


\noindent \textbf{Millicharged DM:} DM carries a small electric charge $\epsilon_\chi e$ and couples to the SM photon, $A_\mu$,
\beq
\mathcal{L} \supset \epsilon_\chi e A_\mu\bar{\chi}\gamma^\mu\chi + A_\mu J^\mu. 
\eeq
$\chi$ particle can either carry a fractional electric charge directly~\cite{Okun:1983vw, Foot:1990uf, Foot:1992ui} or charged under other $U(1)$ which kinetically mixes with the SM hypercharge~\cite{Holdom:1985ag}. $A_\mu$ couples to the SM current, which is given by
\beq
J^\mu=\bar{p}(k^\prime) \left(\frac{e(k+k^\prime)^\mu}{2m_N} + \frac{g_p}{2} \frac{i \sigma^{\mu\nu}q_\nu}{2m_N} \right) p(k) +\bar{n}(k^\prime) \left(\frac{g_n}{2} \frac{i \sigma^{\mu\nu}q_\nu}{2m_N} \right) n(k),
\eeq
where $g_{p}$ and $g_{n}$ are the $g$-factors of protons and neutrons respectively. 

\noindent \textbf{Magnetic dipole DM with a light mediator:} DM does not carry electric charge, but it could still couple to the SM photon through a loop of other charged species. This could generate an anomalous magnetic dipole moment of the DM particle~\cite{Weiner:2012gm}:
\beq\label{eq:magneticDMlight}
\mathcal{L} \supset \frac{i\mu_\chi}{2} \bar{\chi}\sigma^{\mu\nu}\chi F_{\mu\nu} .
\eeq

\noindent \textbf{Magnetic dipole DM with a heavy mediator:} magnetic dipole DM can also interact with the SM electromagnetic current through a heavy mediator. One way is through the kinetic mixing of a broken dark $U(1)_D$ with the SM $U(1)_{\rm em}$
\beq
\mathcal{L} \supset \frac{i\mu_\chi^D}{2} \bar{\chi}\sigma^{\mu\nu}\chi F_{\mu\nu}^D+ \varepsilon F_{ \mu\nu}^D F^{\mu\nu}, 
\eeq
where the dark photon is heavy with a mass larger than the momentum transfer in DD, $m_{D}^{2} \gg q^2$. 
Integrating out the heavy dark photon leads to a coupling between DM and the SM photon as listed in Table~\ref{table:SI_models}. 

\noindent \textbf{Electric dipole DM with a light mediator:} if a dark sector has CP violating interaction, it is also possible that the DM carries an electric dipole moment which aligns with the DM spin. 

\noindent \textbf{Anapole DM with a heavy mediator:} if DM is a majorana fermion, monopole and dipole interactions with electromagnetism is forbidden by CPT symmetry. The only allowed EM coupling is the anapole coupling. For example, the majorana DM could have the anapole moment under a dark broken $U(1)$, which kinetically mixes with the SM photon~\cite{PhysRevD.32.1266}. Integrating out the heavy dark photon, we have 
\beq
\mathcal{L} \supset ig_{\rm ana}\bar{\chi}\gamma^\mu\gamma^5 \chi \partial^\nu F_{\mu\nu}.
\eeq
Early references on anapole DM could be found in~\cite{Pospelov:2000bq, Fitzpatrick:2010br}.

\section{Next-generation experiments} \label{sec:nextgenexp}
\begin{table}[!htb]
\centering
\begin{tabular}{| c | c | c | c |  }
\hline \pbox{10cm}{ \vspace{10pt} Experiment} & \pbox{10cm}{ \vspace{10pt} Target} & Exposure & Energy window  \\[-4pt]
& & $[{\rm ton} \cdot {\rm year}]$ &  $[{\rm keV}]$ \\ \hline
\textsc{Darwin}~\cite{Aalbers:2016jon} & ${}^{131} {\rm Xe}$ & 200 & $[5 - 40]$ \\[3pt]
\textsc{Darkside-20k} (High)~\cite{Aalseth:2017fik}  & ${}^{40} {\rm Ar}$ & 200 & $[32 - 200]$ \\[3pt]
\textsc{Darkside-20k} (Low)~\cite{Agnes:2018ves}  & ${}^{40} {\rm Ar}$ & 200 & $[0.6 - 15]$ \\[3pt]
\hline
\end{tabular}
\caption{Schematic outline of the next-generation experiments included in our results.}\label{tab:futexp}
\end{table}

We describe configurations of the next-generation experiments used in our forecasts in this Appendix. Although there is a rich experimental program underway with different targets, we only focus on experiments that use the dual-phase (liquid-gas) time projection chamber (TPC) technology with noble elements as targets. Table~\ref{tab:futexp} contains a schematic outline of three experiments: a \textsc{Darwin}-like liquid Xenon (LXe) experiment, a \textsc{Darkside-20k}-like liquid Argon (LAr) experiment with both a high and low DM mass program. Each of them are discussed in turn below:

\begin{enumerate}
\item \textsc{Darwin}: The DARk matter WImp search with liquid xenoN (DARWIN) is Generation-3 LXe experiment proposed by the \textsc{Xenon} collaboration. With a 40 ton active volume and 5 year observation time, it's projected 200 ton $\times$ year exposure will allow DM probes to reach the neutrino-floor for DD experiments. We model it based on the conceptual design report~\cite{Aalbers:2016jon} and the latest \textsc{Xenon-1T} configuration. We choose the observation window in recoil energy to be $[5$-$40]$ keV (cf.~\cite{Blanco:2019hah}) and divide it into 19 equally spaced bins for a primary scintillation signal (S1)-only analysis. This assumes near-perfect electron recoil (ER) background subtraction that is ensured by focusing only on the events in the nuclear recoil (NR) region. In practice, we achieve this by convolving our theory recoil spectra with the efficiency curve from the latest \textsc{Xenon} DM analysis (given in Fig. 1 of ref.~\cite{Aprile:2018dbl}) and multiplying by a factor of $0.5$. Following ref.~\cite{Edwards:2017kqw}, we conservatively adopt the rate for NR background components (including a 10\% uncertainty) from ref.~\cite{Aprile:2017iyp}, rescaling them with the appropriate exposure factors. The precise choice of the background should not affect our results, since all our constraints are derived for the signal-limited region with $\mathcal{O}(100)$ total events.  

\item \textsc{Darkside-20k} (High): We follow the official proposal for a neutrino-floor LAr experiment with an integrated exposure of 200 ton$\times$year achieved over a 10 year observation period. Our model detector has an S1-only search region with 19 linearly spaced bins in the range $[32$-$200]$ keV that provides us a unique probe of $\mathcal{O}(0.1$-$1)$ TeV DM candidates. An advantage of using Ar targets is their superior pulse-shape discrimination (PSD) of the S1 signal allowing for nearly background-free detection of any DM events. Thus, we only consider a background rate of 0.1 events over the entire observation period along with a 10\% systematic uncertainty. Finally, we adopt the efficiency curve for NR detection from Fig. 6 of ref.~\cite{Agnes:2014bvk}.

\item \textsc{Darkside-20k} (Low): Although there is no outlined for a low DM mass search using the Gen3 \textsc{Darkside-20k} setup discussed above, we make the science case for one in Sec.~\ref{sec:targets} as a complementary probe to the \textsc{Darwin} experiment. We borrow the configuration in ref.~\cite{Agnes:2018ves} that used the \textsc{Darkside-50} apparatus to perform an ionization (S2)-only analysis, albeit with a 200 ton$\times$year exposure. There are two major differences compared to the high DM mass search: i) an S2-only analysis allows a far lower recoil energy threshold, ii) the PSD is no longer available and we need to contend with higher background rates. Thus, our fiducial analysis has a recoil energy observation window of $[0.6$-$15]$ keV and an optimistic background rate of 1 event for the entire observation period. We also use a constant acceptance of 0.43 following the discussion below Fig.1 of ref.~\cite{Agnes:2018ves}.

\end{enumerate}

%

\bibliography{ref2}
\bibliographystyle{jhep}

\end{document}